\documentclass[12pt,a4paper,final]{iopart}

\usepackage{iopams}  
\expandafter\let\csname equation*\endcsname\relax
\expandafter\let\csname endequation*\endcsname\relax
\usepackage{amsmath}
\usepackage{graphicx}
\usepackage[breaklinks=true,colorlinks=true,linkcolor=blue,urlcolor=blue,citecolor=blue]{hyperref}

\newcommand{\D}{\mathrm{d}}
\newcommand{\half}{\frac{1}{2}}

\newcommand{\vecb}{{\boldsymbol b}}
\newcommand{\vecc}{{\boldsymbol c}}
\newcommand{\vecr}{{\boldsymbol r}}
\newcommand{\vecp}{{\boldsymbol p}}
\newcommand{\vech}{{\boldsymbol h}}

\newcommand{\vecu}{{\boldsymbol u}}
\newcommand{\vecs}{{\boldsymbol s}}
\newcommand{\vecn}{{\boldsymbol \nabla}}
\newcommand{\C}{{\boldsymbol C}}
\newcommand{\vecl}{{\boldsymbol \ell}}

\newcommand{\kbt}{k_{\mathrm{B}}T}

\begin{document}

\title[Chiral active matter]{Chiral active matter: microscopic `torque dipoles' have more than one hydrodynamic description}

\author[cor1]{Tomer Markovich$^{1,2}$, Elsen Tjhung$^1$ and Michael E. Cates$^1$}
\address{$^1$DAMTP, Centre for Mathematical Sciences, University of Cambridge, Cambridge CB3 0WA, United Kingdom}
\address{$^2$Center for Theoretical Biological Physics, Rice University, Houston, TX 77030, USA}
\eads{\mailto{tm36@rice.edu}}

\begin{abstract}
Many biological systems, such as bacterial suspensions and actomyosin networks, form polar liquid crystals.
These systems are  `active' or far-from-equilibrium, due to local forcing of the solvent by the constituent particles.
In many cases the source of activity is chiral; since forcing is internally generated, some sort of  `torque dipole' is then present locally. But it is not obvious how `torque dipoles' should be encoded in the hydrodynamic equations that describe the system at continuum level: different authors have arrived at contradictory conclusions on this issue.
In this work, we resolve the paradox by presenting a careful derivation, from linear irreversible thermodynamics, of the general equations of motion of a single-component chiral active fluid with spin degrees of freedom. We find that there is no unique hydrodynamic description for such a fluid in the presence of torque dipoles of a given strength. Instead, at least three different hydrodynamic descriptions emerge, depending on 
whether we decompose each torque dipole as two point torques, two force pairs, or one point torque and one force pair -- where point torques create internal angular momenta of the chiral bodies (spin), whereas force pairs impart centre of mass motion that contributes to fluid velocity. By considering a general expansion of the Onsager coefficients, we also derive a new shear-elongation parameter and cross-coupling viscosity,
which can lead to unpredicted phenomena even in passive polar liquid crystals.
Finally, elimination of the angular variables gives an effective polar hydrodynamics
with renormalized active stresses, viscosities and kinetic coefficients. Remarkably, this can include a direct contribution of chiral activity to the equation of motion for the polar order parameter, which survives even in `dry' active systems where the fluid velocity is set to zero.
\end{abstract}

\vspace{2pc}

\submitto{\emph{New J. Phys.}}

\section{Introduction}


Active matter is a class of non-equilibrium systems, where energy is injected to the system continuously by the constituent particles themselves~\cite{marchetti}.
Some examples of active fluids are bacterial suspensions, biological tissues and the actomyosin network inside eukaryotic cells.
In the case of bacterial suspensions, the flagella of the bacteria continuously stir the solvent, driving the system out-of-equilibrium.
In addition, the bacteria also tend to align locally with a common tail-to-head direction, forming \emph{polar} rather than \emph{nematic} liquid-crystalline ordering~\cite{wioland,cisneros}.
Similarly, actomyosin network are also known to form polar order during cell motility, in particular,
actin polymerisation at the front drives this motility~\cite{verkhovsky,ziebert,elsen2015}.
On the other hand, some other active fluids such as tissues can also show \emph{nematic} ordering~\cite{saw}.

At a coarse-grained level, the forcing of the solvent by a single bacterium is usually represented as a force dipole.
Upon hydrodynamic averaging, a collection of force dipoles give rise to a symmetric active stress, which drives the system out-of-equilibrium in steady state.
Above certain activity threshold, this active stress may give rise to spontaneous flow transitions~\cite{voituriez}.
However, more recently, it has been suggested that the source of activity may also be \emph{chiral}~\cite{naganathan}.
For instance, in the experiment of~\cite{zhou}, it has been observed that when the bacteria are dead, they are aligned uniformly throughout the system,
but when they are alive, the `polarization' (alignment vector) spontaneously acquires a helical twist.
This suggests that \emph{chiral} active terms should also be present in the hydrodynamic description of active fluids 
-- in addition to the achiral symmetric active stress from the force dipoles.

Different hydrodynamic descriptions for chiral active fluids have been suggested in the literature.
For instance, following ~\cite{elsen2017}, one can add, phenomenologically, a lowest order chiral active stress in the Navier-Stokes equation 
without any other modifications to the dynamics of the polarization field.
However, following ~\cite{furthauer}, one can also add a chiral active term in the equation for conservation of angular momentum, 
which was absent in~\cite{elsen2017}. In both cases, the ultimate source of chiral activity can be considered as a pair of equal and opposite torques or `torque dipole' of a certain strength -- because any locally unbalanced torque would require external rather than internal forcing. Nonetheless, these choices give different hydrodynamic descriptions and predictions, creating significant ambiguity about how chiral activity should be treated at a coarse-grained level.

In order to understand this better, in this paper we present a general formalism for a single-component polar active fluid with explicit angular degrees of freedom. (We will eliminate these carefully later on.)
We derive the hydrodynamic equations for the polarization (alignment) field, linear momentum and angular momentum using linear irreversible thermodynamics.
In contrast to the previous literature, we also go further by expanding the Onsager coefficients in terms of the structural order parameter (the polarization) beyond the linear order.
Interestingly, we thereby derive a new shear-elongation parameter, which allows the magnitude of the polarization  to change under shear flow: the particles become more aligned with increasing shear rate.
This has interesting repercussions, even for passive polar liquid crystals, such as a shear-induced first order phase transition which we studied previously by simply postulating the existence of the shear-elongation parameter~\cite{shear}.
By expanding the dissipative Onsager coefficients, we also derived a cross-coupling viscosity, in addition to shear and rotational viscosity, whose physical role is to couple the rotational degree of freedom to shear.  

Having established the general formalism, we then consider the simplest form of chiral activity, whereby
we treat each active particle as a `torque dipole'.
In the case of bacterial suspensions like \emph{E. coli}, the flagella rotates anti-clockwise whereas the body rotates clockwise.
This indeed looks like a `torque dipole' acting on the fluid. 
However at the mesoscopic level, this `torque dipole' has to be defined more precisely.
First, we represent a `torque dipole' as two equal and opposite point torques, separated by some distance.
Second, we can also represent the same `torque dipole' as two force pairs which are separated by the same distance.
Each force pair acts like a torque and becomes a point torque in some limit.
We discover that these two representations of the same torque dipole (in the limiting case) will in fact give rise to two very different hydrodynamic descriptions,
corresponding to the proposals made in \cite{furthauer} and in \cite{elsen2017} respectively. We also consider a third combination where there is one point torque and one force pair, giving results different from either of these choices.

Finally, we discuss how to eliminate the angular momentum equation to get an effective hydrodynamic description
for just the polarization field and the linear momentum.
We find that the effective hydrodynamic equations have similar form as those of polar active fluids~\cite{elsen2018},
but with renormalized viscosities, kinetic coefficients, and active stresses.
The renormalization of active stresses can becrucial for the stability of the fluid.
Importantly, our effective hydrodynamic equation for the relaxational dynamics polar field may be directly affected by chiral activity without the need for advection.  This opens a new pathway to study `dry' active matter such as bacteria on a substrate or within a gel matrix.
To complete the presentation we generalize these equations to include noise, chosen to satisfy detailed balance in the equilibrium limit.

The outline of this paper is as follows.
In Section~\ref{general-formalism}, we present the formalism to derive our general equations of motion
for a one-component polar liquid crystal with spin degree of freedom.
Then in Section~\ref{active}, we discuss three different representations of a `torque dipole' which will give completely different hydrodynamic descriptions.
In Section~\ref{elimination}, we show how to eliminate the angular momentum degree of freedom carefully and \
obtain renormalized viscosities, kinetic coefficients and active stresses.
Finally in Section~\ref{noise}, we add noise to the equations of motion (after elimination of angular momentum)  
and discuss the so-called `spurious drift' term that must be added to make sense of the noisy equations.
We conclude our findings in Section~\ref{conclusion}.

\section{General Formalism \label{general-formalism}} 

We assume the fluid is made up in three dimensions of identical rigid body particles with mass $m$ and centre-of-mass (CM) position $\vecr_{i}(t)$, where $i$ labels the particle's index. 
Let us also denote the orientation of particle $i$ by a unit vector $\hat{\vecp}_{i}(t)$, called polarity.
For example, in the case of bacteria, $\hat{\vecp}_{i}(t)$ is a body-fixed unit vector which points from the tail to head. 
Finally, we also define the spin angular momentum of particle $i$ to be $\boldsymbol{\ell}^s_i(t)$.
The spin angular momentum describes the rotation of particle $i$ around its CM position $\vecr_i(t)$.

At the hydrodynamic level of description, we define the number density, polarization, hydrodynamic velocity,
and internal angular momentum to be the coarse-grained average of these microscopic variables:
\begin{eqnarray}
n(\vecr,t) 			   	 & = \left\langle \sum_{i}\delta\left(\vecr-\vecr_{i}(t)\right)\right\rangle =\text{constant} \, , \\
\vecp(\vecr,t) 			 & = \frac{1}{n}\left\langle \sum_{i}\hat{\vecp}_{i}(t)\delta\left(\vecr-\vecr_{i}(t)\right)\right\rangle \label{eq:p-micro} \, , \\
\rho\boldsymbol{u}(\vecr,t) & = \left\langle \sum_{i}m\boldsymbol{v}_{i}(t)\delta\left(\vecr-\vecr_{i}(t)\right)\right\rangle \, , \\
\boldsymbol{\ell}(\vecr,t)     & = \left\langle \sum_{i}\boldsymbol{\ell}^s_{i}(t)\delta\left(\vecr-\vecr_{i}(t)\right)\right\rangle \, , \label{eq:ell-micro}
\end{eqnarray}
where $\boldsymbol{v}_{i}(t)=\D\vecr_{i}/\D t$ is the velocity of particle $i$. 
For simplicity, we assume the number density to be constant everywhere in space. 
Consequently, the mass density $\rho=nm$ is also constant and the conservation of mass dictates that $\vecn\cdot\vecu=0$ (incompressibility). 
The generalisation of these results to compressible fluid is given in~\ref{app-inertia}. 

\begin{figure}
\centering
\includegraphics[width=0.9\textwidth]{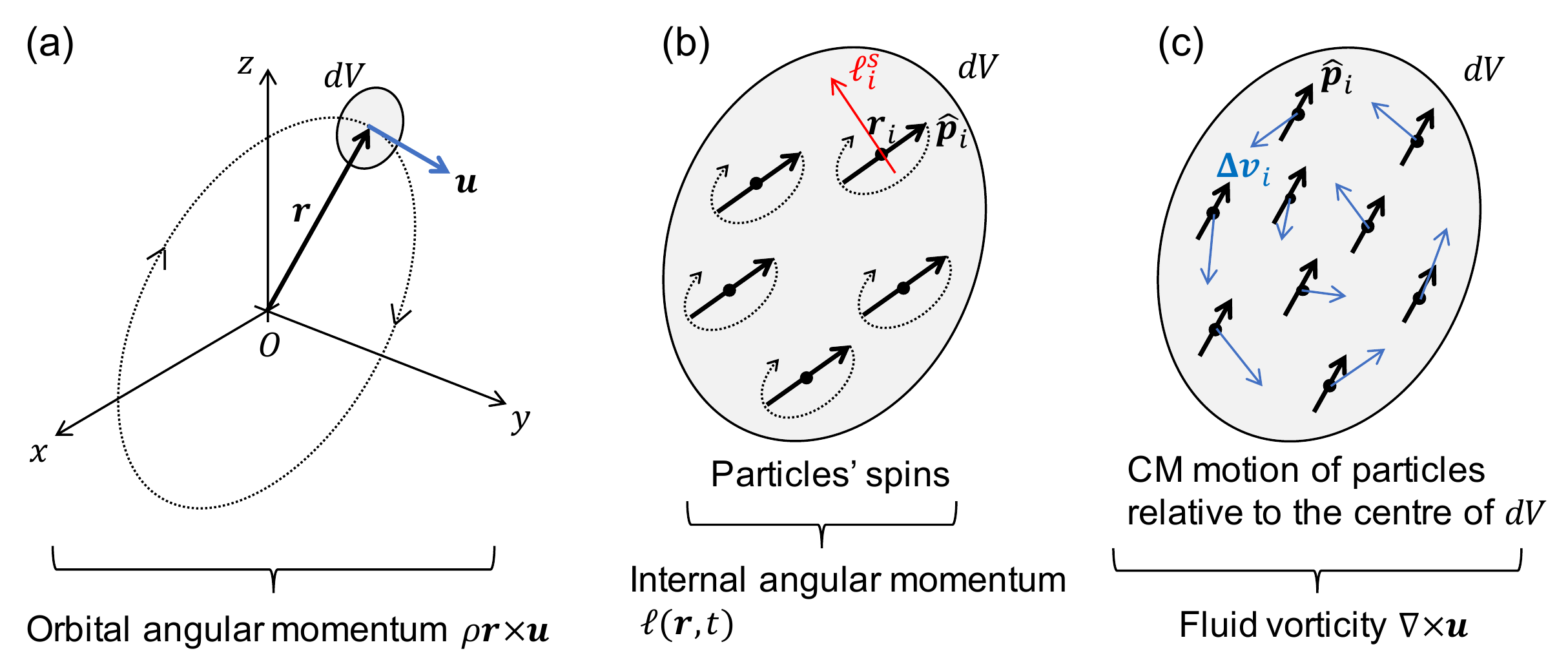}
\caption{
(a) $\rho\vecr\boldsymbol{\times u}$ is the orbital angular momentum of a volume element $\D V$ about the origin. 
(b) The internal angular momentum $\boldsymbol{\ell}(\vecr,t)$ consists of particles' spins, $\boldsymbol{\ell}_{i}^s$ inside $\D V$.
(c) The fluid vorticity $\nabla\times\boldsymbol{u}$ describes the CM rotations of the particles around the hydrodynamic centre $\vecr$. \label{fig:angular-momentum}}
\end{figure}

In \eqref{eq:ell-micro}, $\boldsymbol{\ell}(\vecr,t)$ is the \emph{internal} angular momentum, 
defined here to be the average of all particles' spins inside some mesoscopic volume $\D V$ located at $\vecr$, 
see Fig.~\ref{fig:angular-momentum}(b)~\cite{lubensky2005}.
In general, this need not be the only contribution:
$\boldsymbol{\ell}(\vecr,t)$ may also contain the particles' angular momenta relative to the hydrodynamic centre, $m(\vecr_{i}-\vecr)\times(\boldsymbol{v}_{i}-\vecu)$, 
see Fig.~\ref{fig:angular-momentum}(c). This contribution has been discussed in~\cite{klymko} and would arise, for instance, in a system of active particles that swim on circular orbits whose radius is small compared to a coarse-grained fluid volume element (creating an angular momentum density unrelated to either particle spin or fluid vorticity). 
In this paper, however, we assume this sub-orbital motion of the particles around the hydrodynamic centre $\vecr$ to be the same as the fluid vorticity $\vecn\times\vecu$ [Fig.~\ref{fig:angular-momentum}(c)] 
and thus in (\ref{eq:ell-micro})
$\boldsymbol{\ell}(\vecr,t)$ encodes spin dynamics only.

We also define the nematic order parameter to be:
\begin{equation}
Q_{\alpha\beta}(\vecr,t) = \frac{1}{n}\left\langle \sum_{i}\left(\hat{p}^i_{\alpha}(t)\hat{p}^i_{\beta}(t)-\frac{\delta_{\alpha\beta}}{3}\right)\delta\left(\vecr-\vecr_{i}(t)\right)\right\rangle .\label{eq:Q-micro}
\end{equation}
In the hydrodynamic theory, $\boldsymbol{Q}$ frequently appears, for instance as the source of active stress; in principle it can vary independently of $\vecp$. However, like most authors, we aim for a simplified description of polar liquid crystals in which $\vecp$ is the only orientational degree of freedom retained at hydrodynamic level. To achieve this, it is quite common to make the following approximation
(see~\cite{elsen2018}):
\begin{equation}
Q_{\alpha\beta}\simeq p_{\alpha}p_{\beta}-\left|\vecp\right|^{2}\frac{\delta_{\alpha\beta}}{3}, \label{eq:Q-approx}
\end{equation}
with $|\vecp|^{2}$ and $\vecp$ setting the principal eigenvalue and eigenvector of $\boldsymbol{Q}$, respectively. Formally this is a strong alignment approximation, becoming exact when $\hat{\vecp}_{i}$ is the same for all particles $i$. We will use it below whenever an expression for $\boldsymbol{Q}(\vecp)$ is needed.

\subsection{Conservation laws}

The rate of change of the linear momentum within some volume element is given by the Navier-Stokes (Cauchy) equation:
\begin{equation}
\rho\frac{Du_\alpha}{Dt}=\partial_\beta\sigma_{\alpha\beta} + f_\alpha, \label{eq:udot}
\end{equation}
where $\frac{D}{Dt}=\frac{\partial}{\partial t}+\boldsymbol{u}\cdot\vecn$ is the convective derivative.
Here $\boldsymbol{\sigma}$ is the stress tensor, which gives a surface force acting across the boundary of each volume element $\D V$, originating from a short-range two-body interaction between the CM positions of the particles, while $\boldsymbol{f}$ is an external body force, \emph{e.g.}~gravity.  Because we deal with antisymmetric stresses, the order of indices on $\sigma_{\alpha\beta}$ in \eqref{eq:udot} is important; there is no fixed convention in the literature but we use that of \cite{LL} in this paper.

At this point, let us decompose the stress tensor into symmetric and anti-symmetric part: $\sigma_{\alpha\beta}=\sigma_{\alpha\beta}^{s}+\sigma_{\alpha\beta}^{a}$.
Furthermore, let us write the anti-symmetric part of the stress tensor in the following form:
\begin{equation}
\sigma_{\alpha\beta}^{a} = \tilde{\sigma}_{\alpha\beta}^{a}
				      + \partial_{\delta}\left[\frac{1}{2}\epsilon_{\alpha\beta\gamma}C_{\gamma\delta}^{\sigma}\right], \label{eq:sigmaa}
\end{equation}
where $\tilde{\sigma}_{\alpha\beta}^{a}$ cannot be written as the divergence of any third-rank tensor. In general this decomposition into a divergence term and a remainder is not expected to be unique. However, 
as we shall see later, the divergence part of the anti-symmetric stress, $\C^\sigma$, will contribute to a surface torque, 
whereas the non-divergence part, $\tilde{\boldsymbol{\sigma}}^a$, will contribute to a body torque in the angular momentum equation.
These considerations will allow us to unambiguously separate these two contributions which will thereafter enter the hydrodynamic description in distinct ways.

\begin{figure}
\centering
\includegraphics[width=0.7\textwidth]{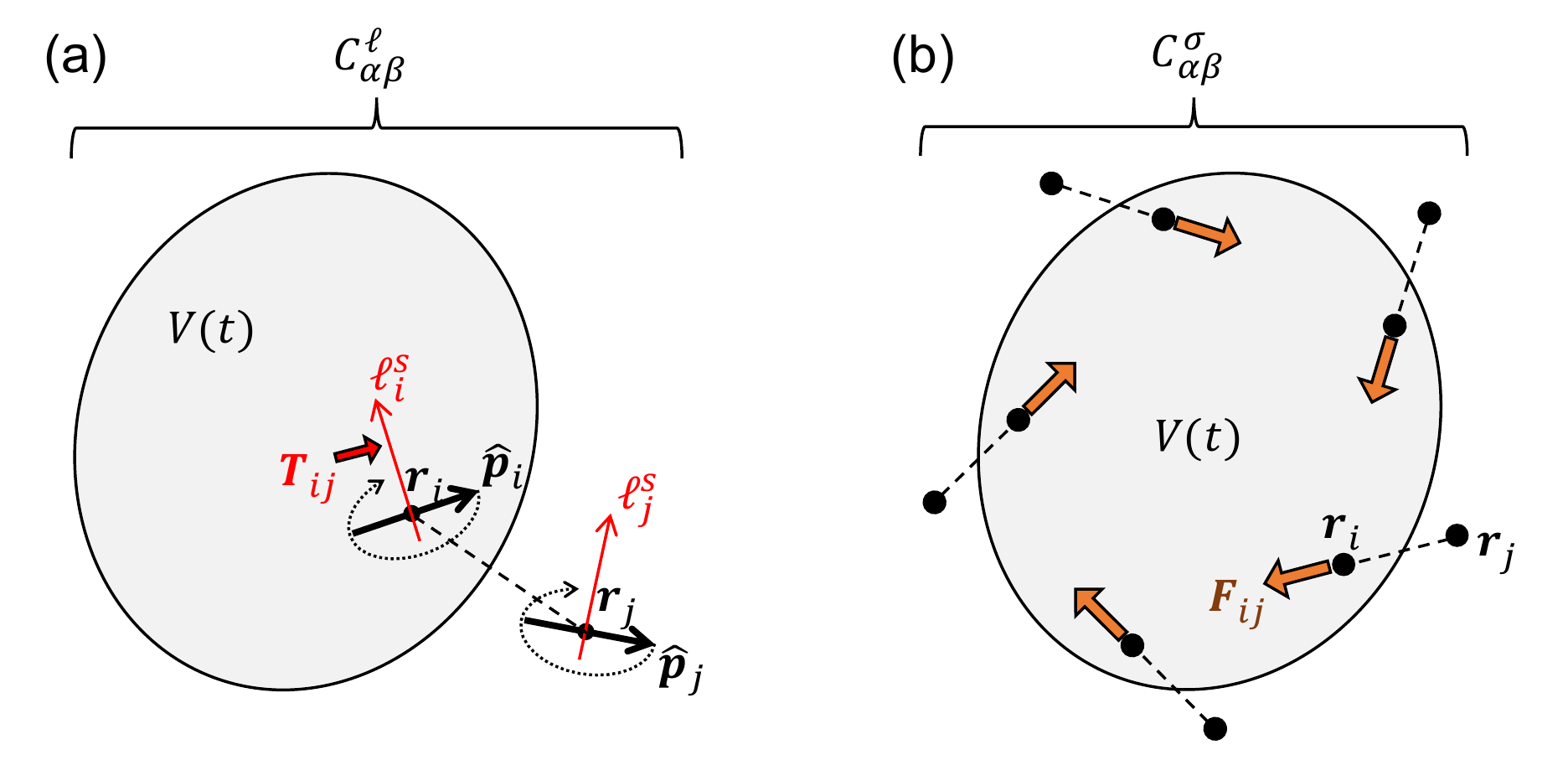}
\caption{There are two contributions to the surface torque:
(a) $C_{\alpha\beta}^{\ell}$, which comes from spin-spin interaction, and 
(b) $C_{\alpha\beta}^{\sigma}$, which comes from two-body interaction.
$\frac{1}{2}\epsilon_{\alpha\beta\gamma}\partial_\delta C_{\gamma\delta}^{\sigma}$ is the divergence part of the anti-symmetric stress tensor $\sigma_{\alpha\beta}^a$, Eq.~(\ref{eq:sigmaa}).
\label{fig:surface-torque}}
\end{figure}

Let us define $\boldsymbol{L}(\vecr,t)$ to be the total angular momentum density of the fluid, 
consisting of orbital and internal angular momentum densities:
\begin{equation}
\boldsymbol{L}(\vecr,t) = \underbrace{\rho\vecr\times\boldsymbol{u}}_{\text{orbital}}
						 + \underbrace{\boldsymbol{\ell}(\vecr,t)}_{\text{intrinsic}}. \label{eq:L}
\end{equation}
The orbital part represents the rotation of the volume element $\D V$ about the origin [see Fig.~\ref{fig:angular-momentum}(a)]
whereas the internal part, $\boldsymbol{\ell}(\vecr,t)$, consists of particles' spins [see Fig.~\ref{fig:angular-momentum}(b)].
The equation of motion for $\boldsymbol{\ell}$ is
\begin{equation}
\frac{D\ell_{\alpha}}{Dt} = \partial_{\beta}C_{\alpha\beta}^\ell + s_{\alpha} + \tau_{\alpha}, \label{eq:elldot}
\end{equation}
where $\C^{\ell}$ is a surface torque that originates from a short-range spin-spin interaction.
For instance, as shown in Fig.~\ref{fig:surface-torque}(a), spin $j$ tends to align with spin $i$ by their exerting short-range torques $\pm\boldsymbol{T}_{ij}$ on each other.
Hence, $\C^{\ell}$ is the analogue of the stress tensor for the angular momentum, and is often referred to as `couple-stress' in the literature. 
In (\ref{eq:elldot}), $\boldsymbol{s}$ and $\boldsymbol{\tau}$ are both body torques. 
Unlike the body force $\boldsymbol{f}$ in (\ref{eq:udot}), which is always external, 
the body torques can be internal ($\boldsymbol{s}$) or external ($\boldsymbol{\tau}$).
Next, we shall derive the relationship between the internal body torque, $\boldsymbol{s}$, and the stress tensor, $\boldsymbol{\sigma}$. 

Consider a co-moving finite parcel of fluid occupying a region $V(t)$. The rate of change of the total angular momentum of this fluid parcel is:
\begin{align}
\nonumber\frac{\D}{\D t}\int_{V(t)}L_{\alpha}\D V & = \int_{V(t)}\left( \rho\,\epsilon_{\alpha\beta\gamma}r_{\beta}\frac{Du_{\gamma}}{Dt} + \frac{D\ell_{\alpha}}{Dt} \right)\D V \\
 						   & = \int_{V(t)}\left( \epsilon_{\alpha\beta\gamma}r_{\beta}\partial_{\delta}\sigma_{\gamma\delta}
						   		+ \epsilon_{\alpha\beta\gamma}r_{\beta}f_{\gamma}
								+ \partial_{\beta}C_{\alpha\beta}^{\ell}
								+ s_{\alpha}
								+ \tau_{\alpha}\right) \D V, \label{eq:Ldot1}
\end{align}
where we have used Reynolds transport theorem in the first line and substituted (\ref{eq:udot}) and (\ref{eq:elldot}) in the second line. 
Using integration by parts and equation (\ref{eq:sigmaa}), this becomes:
\begin{align}
\frac{\D}{\D t}\int_{V(t)}L_{\alpha}\,\D V & = \left[ \oint_{\partial V(t)}\vecr\times\boldsymbol{\sigma}\cdot \D\boldsymbol{S}\right]_{\alpha}
							+ \int_{V(t)}\left[\vecr\times\boldsymbol{f}\right]_{\alpha}\D V 
						        + \oint_{\partial V(t)} \left(C_{\alpha\beta}^{\ell} + C_{\alpha\beta}^{\sigma}\right)\D S_\beta \nonumber\\
 						     & + \int_{V(t)}\left(s_{\alpha}+\epsilon_{\alpha\beta\gamma}\tilde{\sigma}_{\beta\gamma}^{a}\right)\D V
						         + \int_{V(t)}\tau_{\alpha}\,\D V. \label{eq:Ldot2}
\end{align}
The first two terms in (\ref{eq:Ldot2}) give torques about the origin whose causes are the same as for spinless particles. 
The third term in (\ref{eq:Ldot2}) is a surface torque. 
The surface torque consists of $\C^{\ell}$, which comes from spin-spin interactions, 
and also $\C^{\sigma}$, which comes from the two-body interaction between particles' CM positions. 
For example, the configuration of point particles in Fig.~\ref{fig:surface-torque}(b) can give rise to a surface torque \emph{via} this two-body interaction. 
Finally the last two terms in (\ref{eq:Ldot2}) are volume terms. 
By Newton's third law, only the \emph{external} body torque should produce such a volume term in (\ref{eq:Ldot2}); 
all the internal forces and torques should add up to a surface term. 
Thus we must have the following equality 
\begin{equation}
s_{\alpha}=-\epsilon_{\alpha\beta\gamma}\tilde{\sigma}_{\beta\gamma}^{a},
\label{eq:intorq}
\end{equation}
and the internal angular momentum equation (\ref{eq:elldot}) becomes:
\begin{equation}
\frac{D\ell_{\alpha}}{Dt} = \partial_{\beta}C_{\alpha\beta}^\ell - \epsilon_{\alpha\beta\gamma}\tilde{\sigma}_{\beta\gamma}^{a} + \tau_{\alpha}. \label{eq:elldot1}
\end{equation}
Eq.~\eqref{eq:intorq} states that the non-divergence part of the anti-symmetric stress, $\tilde{\boldsymbol{\sigma}}^a$ in (\ref{eq:sigmaa}), can be identified as the \emph{internal} body torque. This makes the decomposition \eqref{eq:sigmaa} physically unambiguous, as was promised above. It requires however that the internal body torque is separately identifiable in any constitutive relations connecting the state of antisymmetric stress with the hydrodynamic variables $\vecp$ and $\vecu$. This in turn may require closer examination of the microscopic physics than expected at that level.

The observation that the divergence part of the anti-symmetric stress, $\C^\sigma$ in (\ref{eq:sigmaa}), 
does not contribute to the internal angular momentum equation (\ref{eq:elldot1})  is already known~\cite{LL,forster}.
In fact, there is a transformation that allows the term in $\C^\sigma$ to be rewritten in terms of a {\em symmetric} stress without changing the equation of motion for {\em linear} momentum~\cite{furthauer,LL}.
This might suggest that $\C^\sigma$ can also be somehow absorbed into the symmetric part of the stress tensor for the purposes of (\ref{eq:elldot1}), in which case this equation  could correctly be written with the full $\boldsymbol{\sigma}^a$ replacing $\tilde{\boldsymbol{\sigma}}^a$ on the right hand side. The resulting equation indeed has been used in some of the recent liquid crystal and active matter literature such as~\cite{furthauer,lubensky2005}. However such a substitution is not generally correct; we shall see below that it gives a wrong hydrodynamic description for a particular type of chiral active particles (Case Ib in Sec.~\ref{CaseI}). Happily this was not the type considered in \cite{furthauer} whose more specific results are therefore unaffected.

Finally we introduce the angular velocity vector $\boldsymbol{\Omega}$, of a volume element to be:
\begin{equation}
\ell_{\alpha}(\vecr,t)=I_{\alpha\beta}(\vecr,t)\Omega_{\beta}(\vecr,t)=I\Omega_{\alpha}(\vecr,t),\label{eq:Omega}
\end{equation}
where $I_{\alpha\beta}(\vecr,t)$ is the (molecular) moment of inertia per unit volume. This depends in general on the polarization $\vecp$ but for simplicity, in the discussion below, we assume its isotropic: $I_{\alpha\beta}=I\delta_{\alpha\beta}$ where, for an incompressible fluid, $I$ is a constant. This arises, {\em e.g.}, for particles whose polarization is defined by their swimming direction but are otherwise spherical.
In~\ref{app-inertia}, we discuss how the results in this paper can be generalized to the case of anisotropic moment of inertia and a compressible fluid. In what follows we generally represent the angular velocity of particles not by a vector $\Omega_\gamma$ but by the rotation rate tensor $-\Omega_{\alpha\beta} = -\epsilon_{\alpha\beta\gamma}\Omega_\gamma$ and do the same for the fluid vorticity ${\omega}_\gamma$. This saves incessant use of the cross product; the symbols $\boldsymbol{\Omega},\boldsymbol{\omega}$ can represent either quantity, depending on context.

\subsection{Equilibrium case}

Assuming local equilibrium, an effective coarse-grained free energy functional can be written as,
\begin{equation}
F[\vecp,\boldsymbol{u},\boldsymbol{\ell}] = \int\left[ \frac{1}{2}\rho u^{2}
									   + \frac{1}{2}\ell_{\alpha}I_{\alpha\beta}^{-1}\ell_{\beta}
									   + \mathbb{F}_{0}(p_{\alpha},\partial_{\alpha}p_{\beta})\right] \D V. \label{eq:F}
\end{equation}
The first and second term in (\ref{eq:F}) are the translational and rotational kinetic energy respectively, whereas 
$F_{0}[\vecp]=\int \mathbb{F}_{0}(\vecp,\vecn\vecp)dV$ is the configurational free energy. 
For instance, $\mathbb{F}_{0}$ might be written as a Landau-Ginzburg expansion in $\vecp$ and $\vecn\vecp$. 
We shall also define the molecular field to be $\boldsymbol{h}=\frac{\delta F_{0}}{\delta\vecp}$
(note that in some literature this is defined with a minus sign),
and introduce the symmetric and anti-symmetric parts of the velocity gradient tensor: $\nu_{\alpha\beta} =\frac{1}{2}\left(\partial_{\alpha}u_{\beta}+\partial_{\beta}u_{\alpha}\right)$,
and $\omega_{\alpha\beta}  =\frac{1}{2}\left(\partial_{\alpha}u_{\beta}-\partial_{\beta}u_{\alpha}\right)$, respectively.
Note that $\boldsymbol{\omega}$ is the fluid vorticity, which involves rotations of CM positions of the particles around the hydrodynamic centre $\vecr$ of the volume element $\D V$
[Fig.~\ref{fig:angular-momentum}(c)].
This is different from $\boldsymbol{\Omega}$ in (\ref{eq:Omega}), 
which describes particles within the volume element spinning around their CM positions [Fig.~\ref{fig:angular-momentum}(b)].

The equation of motion for $\vecp$ can then be derived phenomenologically as in~\ref{app-expansion} yielding,  
\begin{equation}
\frac{\partial p_{\alpha}}{\partial t} + (\boldsymbol{u}\cdot\vecn)p_{\alpha}+\Omega_{\alpha\beta}p_{\beta} = 
		- \gamma_{\alpha\beta} h_{\beta}
		+ \xi_{1}(\Omega_{\alpha\beta}-\omega_{\alpha\beta})p_{\beta}
		+ \xi_{0}\nu_{\alpha\beta}p_{\beta}
		+ \xi_{2}(\vecp\cdot\boldsymbol{\nu}\cdot\vecp)p_{\alpha}. \label{eq:pdot}
\end{equation}
Here the terms on the left hand side are all required by Galilean and rotational invariance. (On elimination of spin degrees of freedom $\Omega_{\alpha\beta}$ will become $\omega_{\alpha\beta}$ whose appearance in this context may be more familiar \cite{elsen2018}.)
The first term on the right hand side is the relaxation term which tends to minimize the free energy, with $\gamma_{\alpha\beta}$ being the (positive) relaxation rate; 
$\xi_0$ is the shear-aligning parameter, which controls whether $\vecp$ aligns or tumbles under an imposed shear flow, and
$\xi_2$ is the shear-elongation parameter.
This term can be obtained from our systematic expansion of the Onsager coefficients (see~\ref{app-expansion}) or 
through coarse-graining of some specific microscopic models~\cite{kung}.
Physically this term can stretch (or compress) the magnitude of $\vecp$ under an imposed shear flow: the particles become more aligned (or more disordered) as we increase the shear rate. 
Some physical consequences of this term were presented by us in~\cite{shear}.
Finally, $\xi_1$ in (\ref{eq:pdot}) is a rotational alignment parameter, analogous to $\xi_0$. 

Equation (\ref{eq:pdot}) should be solved together with the Navier-Stokes/Cauchy equation (\ref{eq:udot}) and the angular momentum equation (\ref{eq:elldot}).
They are coupled together \emph{via} the stress tensor $\boldsymbol{\sigma}$ and couple-stress $\boldsymbol{C}^{\ell}$,
which are both derived in~\ref{app-expansion}. 

The stress tensor consists of dissipative ($\boldsymbol{\sigma}^{d}$), elastic ($\boldsymbol{\sigma}^{e}$), and reactive parts ($\boldsymbol{\sigma}^{r}$). 
The dissipative part is responsible for heat dissipation to the environment and hence contributes to the total rate of entropy production. 
(Note that the $\boldsymbol{\gamma}$-term in (\ref{eq:pdot}) is also a dissipative term.) 
In general, $\boldsymbol{\sigma}^{d}$ can be written as:
\begin{align}
\sigma_{\alpha\beta}^{d} &= \eta_{\alpha\beta\gamma\delta}\nu_{\gamma\delta} 
				      + \eta'_{\alpha\beta\gamma\delta}(\Omega_{\gamma\delta}-\omega_{\gamma\delta}) \nonumber\\
				   &+ \frac{\eta_c}{2}[(\Omega_{\alpha\mu}-\omega_{\alpha\mu})p_\mu p_\beta + (\Omega_{\beta\mu}-\omega_{\beta\mu})p_\mu p_\alpha] 
				      + \frac{\eta_c}{2}(\nu_{\alpha\mu}p_\mu p_\beta - \nu_{\beta\mu}p_\mu p_\alpha) , \label{eq:sigmad}
\end{align}
where $\eta_{\alpha\beta\gamma\delta}$ is the shear viscosity and $\eta'_{\alpha\beta\gamma\delta}$ is the rotational viscosity.
Note that the shear viscosity is generally different in the direction parallel or perpendicular to $\vecp$ (similarly for rotational viscosity).
Thus, in general, $\boldsymbol{\eta}$ and $\boldsymbol{\eta}'$ are functions of $\vecp$, which are given in~\ref{app-expansion}.
The role of the rotational viscosity $\boldsymbol{\eta}'$ is to relax the particles' spins towards $\Omega_{\alpha\beta}=\omega_{\alpha\beta}$.
Thus, rotational dissipation vanishes when the average spin of the particles is equal to the vorticity of the fluid at that particular point. 
Finally, $\eta_c$ is a cross-viscosity, which couples shear to spin degree of freedom (see also \cite{lubensky2005}). 
This comes from the cross-coupling in the Onsager matrix, and can be shown to be dissipative (see~\ref{app-expansion}).
The $\eta_c$-term is interesting because it will renormalize the shear viscosity and the active stress when we eliminates the angular degree of freedom in Section~\ref{elimination} below.

The reactive stress, $\boldsymbol{\sigma}^{r}$, has the form:
\begin{align}
\sigma_{\alpha\beta}^{r} & = -\frac{\xi_1}{2}(p_{\alpha}h_{\beta} - p_{\beta}h_{\alpha})
+ \frac{\xi_0}{2}(p_{\alpha}h_{\beta} + p_{\beta}h_{\alpha})
+ \xi_2 (\vecp\cdot\boldsymbol{h})p_\alpha p_\beta \, . \label{eq:sigmar}
\end{align}
The first term above is anti-symmetric whereas the rest are symmetric with respect to $\alpha\leftrightarrow\beta$.
These reactive terms are identifiable by their dependence on the coefficients $\xi_{0,1,2}$, which  give contributions in \eqref{eq:pdot} that couple $\vecp$ to quantities that change sign on time reversal. As a result, these terms do not contribute to entropy production \cite{chaikin}. Partly for this reason, they are sometimes treated as contributions to the elastic stress and can be derived as such by considering how $\vecp$ responds non-affinely to an instantaneous deformation \cite{elsen2018}. 

For current purposes, however, we reserve the term `elastic stress' (or Ericksen stress) for the stress $\boldsymbol{\sigma}^{e}$ arising from a purely  affine elastic deformation of the material; this does not involve the kinetic coefficients $\xi_{0,1,2}$. Like the reactive stress, it does not contribute to entropy production.
The form of $\boldsymbol{\sigma}^{e}$ is derived in~\ref{app-expansion} and also in~\cite{furthauer,kruse}, as
\begin{equation}
\sigma_{\alpha\beta}^{e} = (\mathbb{F}_{0}-\vecp\cdot\boldsymbol{h})\delta_{\alpha\beta}
				       - \frac{\partial \mathbb{F}_{0}}{\partial(\partial_{\beta}p_{\gamma})}(\partial_{\alpha}p_{\gamma})\, . \label{eq:sigmae}
\end{equation}
The corresponding Gibbs-Duhem relation is $\partial_\beta\sigma_{\alpha\beta}^e = - p_\beta\partial_\alpha h_\beta$ which, used in the Navier-Stokes equation, allows the elastic term there to be viewed as an effective body force if desired (at the cost of disguising its status as a total derivative).
%
The anti-symmetric part of $\boldsymbol{\sigma}^{e}$ is:
\begin{equation}
\sigma_{\alpha\beta}^{a,e} = \underbrace{\frac{1}{2}\left(p_{\alpha}h_{\beta}-p_{\beta}h_{\alpha}\right)}_{\tilde{\sigma}_{\alpha\beta}^{a,e}}
					+ \partial_{\gamma} \Bigg[ \frac{1}{2} 
					           	\underbrace{\left( \frac{\partial \mathbb{F}_{0}}{\partial(\partial_{\gamma}p_{\beta})}p_{\alpha}
									         - \frac{\partial \mathbb{F}_{0}}{\partial(\partial_{\gamma}p_{\alpha})}p_{\beta} \right)}_{\epsilon_{\alpha\beta\pi}C_{\pi\gamma}^{\sigma,e}}
								       \Bigg], \label{eq:sigmaae}
\end{equation}
where we have decomposed $\boldsymbol{\sigma}^{a,e}$ into a divergence part and a remainder, consistent with \eqref{eq:sigmaa}. This decomposition is also consistent with \eqref{eq:intorq} if we interpret the first term as the body torque density arising from the internal molecular field. In principle, however, the second term might include additional body torque contributions whose functional dependence on $\vecp$ happened to be of the divergence form given. Knowledge of the free energy functional $F_0$ alone may not be enough to decide this question, in which case we are making the simplest choice consistent with the chosen $F_0[\vecp]$.

Finally, $C_{\alpha\beta}^\ell$, and also the non-elastic part of $C_{\alpha\beta}^\sigma$,  are proportional to $\partial_\beta\Omega_\alpha$ (see~\ref{app-expansion}).
These will give higher order gradient terms $\sim\vecn\vecn\boldsymbol{u}$ after elimination of angular momentum (see Section \ref{elimination} below).
Thus in the passive case considered so far, we can set $C_{\alpha\beta}^\ell=0$. It follows that 
 in the active case, which we address next, we can take $C_{\alpha\beta}^\ell=C_{\alpha\beta}^{\ell,A}$ to be a purely active contribution.

\section{Adding activity \label{active}} 

Consider now the case in which the particles are active. 
For example, for bacteria, the flagella on its surface exerts
a distribution of point forces $\boldsymbol{F}$ and point torques $\boldsymbol{T}$ on the surrounding solvent. (Likewise cilia on ciliated micro-organisms.) 
A defining feature of active matter is that the net force and the net torque exerted by each particle on the fluid has to be zero:
\begin{align}
\sum\boldsymbol{F} &= \boldsymbol{0} \label{eq:F-balance} \, , \\
\sum\left(\boldsymbol{T} + \delta\vecr_i\times\boldsymbol{F} \right) &= \boldsymbol{0} , \label{eq:T-balance}
\end{align}
otherwise, the particle is no longer active but rather driven by an external force/torque. 
Note that in the torque balance condition above, $\boldsymbol{T}$ is a point torque and 
$\delta\vecr_i\times\boldsymbol{F}$ is the torque created by a point force $\boldsymbol{F}$ on the CM of particle $i$.
Furthermore, $\boldsymbol{F}$ and $\boldsymbol{T}$ can be oscillating in time but here we take them to be time-averaged quantities.

\begin{figure}
	\begin{centering}
		\includegraphics[width=0.9\textwidth]{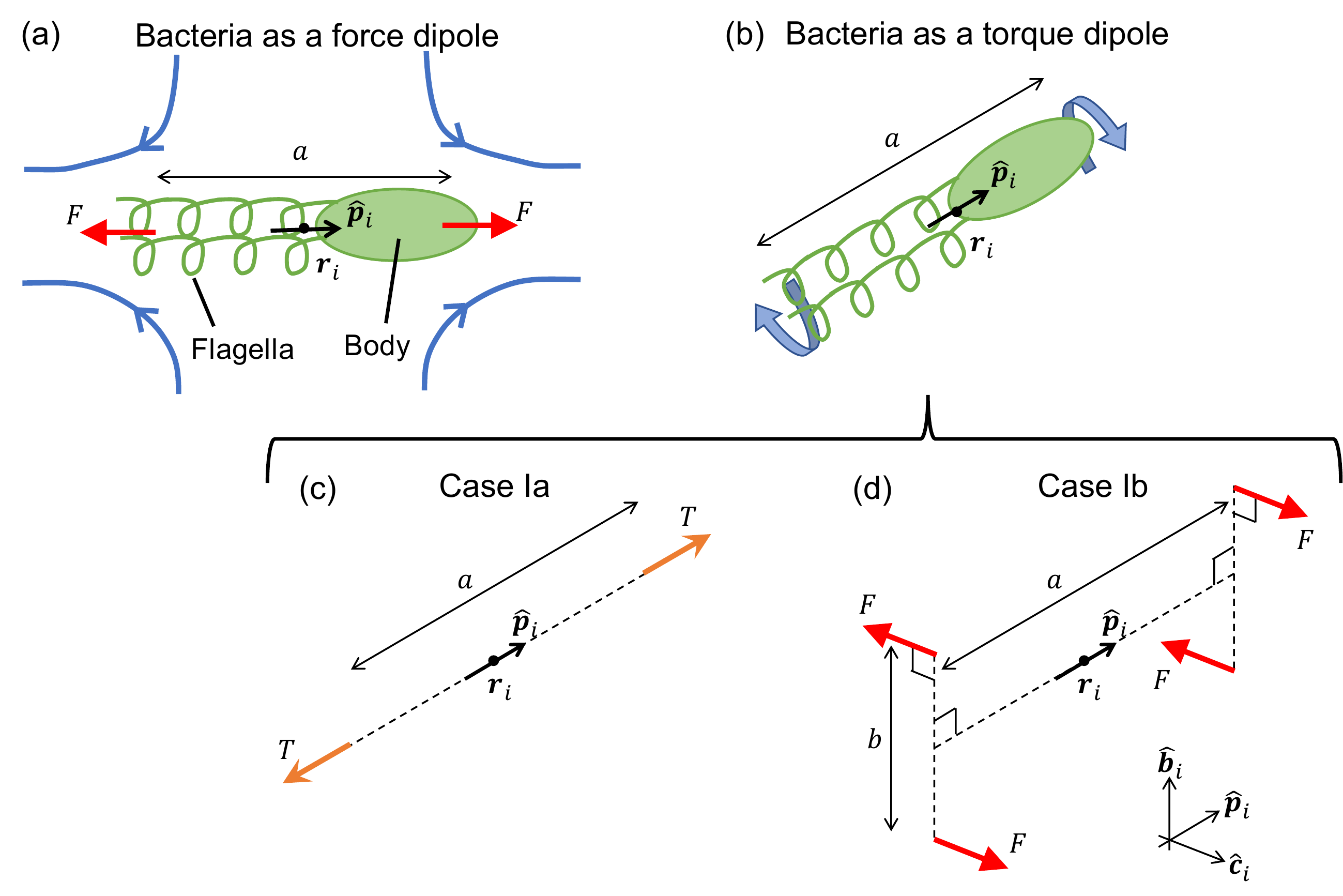}
		\par\end{centering}
	\caption{
		(a) A bacterium as a force dipole which pushes the fluid away along its axis $\hat{\vecp}_i$.
		(b) The bacterium also creates counter-rotating fluid flow, which can be approximated as a `torque dipole'.
		(c,d) Different representations of the same `torque dipole' give rise to different hydrodynamic descriptions. \label{fig:bacteria} }
\end{figure}

We assume that all the passive contributions to the dynamical equations are captured by the passive theory just described.
The presence of these active forces and torques will give an additional active stress tensor, $\boldsymbol{\sigma}^A$, a couple-stress, $\boldsymbol{C}^{\ell,A}$, and an internal body torque, $\boldsymbol{s}^A$, in the hydrodynamic equations:
\begin{align}
\rho\frac{D\vecu}{Dt} &= \vecn\cdot(\boldsymbol{\sigma}+\boldsymbol{\sigma}^{A})  \label{eq:udot2} \, ,\\
\frac{D\vecl}{Dt}         &=  \vecn\cdot\C^{\ell,A} + \vecs +  \vecs^{A} \, . \label{eq:elldot2}
\end{align}
Here and below, we only consider active but otherwise unforced systems, so 
we have set the \emph{external} body force $\boldsymbol{f}$ and the \emph{external} body torque $\boldsymbol{\tau}$ to zero in the equations above. (Such exterior forcing could easily be restored if desired.) The superscripts `$A$' indicate the non-equilibrium or active terms.
The terms without the superscripts `$A$' are the equilibrium or passive terms given in (\ref{eq:sigmad}-\ref{eq:sigmar}). Note that, as explained above, there is no passive contribution to $\boldsymbol{C}^{\ell}$ for the theory chosen here. 
In general, a distribution of point forces on each particle will contribute to $\boldsymbol{\sigma}^{A}$ 
whereas a distribution of point torques will contribute to $\boldsymbol{C}^{\ell,A}$ and $\boldsymbol{s}^{A}$.
Together with (\ref{eq:udot2}-\ref{eq:elldot2}), we must also have the following equalities for the internal body torques:
\begin{align}
s_{\alpha} & =-\epsilon_{\alpha\beta\gamma}\tilde{\sigma}_{\beta\gamma}^{a} \, ,\label{eq:sA0}\\
s_{\alpha}^{A} & =-\epsilon_{\alpha\beta\gamma}\tilde{\sigma}_{\beta\gamma}^{a,A}\, . \label{eq:sA}
\end{align}
Note that we take the condition \eqref{eq:intorq} on the total internal torque to apply separately, as now written in (\ref{eq:sA0},\ref{eq:sA}), for the passive and active parts.
In particular, the condition on $\boldsymbol{s}^{A}$ is equivalent to the requirement of a vanishing sum of active torques (\ref{eq:T-balance}).

An additional and somewhat separate effect of activity is to allow a term which breaks Galilean invariance $\propto(\vecp\cdot\vecn)\vecp$ in (\ref{eq:pdot}).
This active term describes particles' swimming in the direction of $\vecp$ (self-advection). 
The physical significance of this term is described in~\cite{EPR}. 
We shall ignore this self-advection term in the derivations below but note that it is not changed by the elimination of the angular momentum variable. Therefore it can be added back to the final hydrodynamic equations if desired.

The simplest form of a force distribution on each active particle is a permanent force dipole. 
For example, in the case of bacteria, the rotation of the flagella and the counter-rotation of the body tend to expel the fluid away along the direction of $\pm\hat{\vecp}_i$, drawing it in equatorially
[see Fig.~\ref{fig:bacteria}(a)].
This type of particle is called a pusher or extensile, and can be approximated as a force dipole $\pm\boldsymbol{F}$ as shown in the figure. (Reversing the signs gives a puller or contractile particle.)
Upon coarse-graining, the force dipoles will give rise to an active stress $\sigma_{\alpha\beta}^{A}=-Fan\,p_{\alpha}p_{\beta}$~\cite{hatwalne},  
where $a$ is a length governed by the size of the bacteria and $n$ is the number density.
(For pullers/contractile, the sign of the active stress is positive.)
This form of the active stress alone was shown to give rise to interesting phenomena such as spontaneous bulk flow transitions~\cite{voituriez}, 
negative viscosity~\cite{shear,giomi}, and spontaneous droplet motility~\cite{elsen2012}. 

The active stress arising from such a force dipole is achiral. 
However in many biological systems, the source of activity (be it $\boldsymbol{\sigma}^{A}$, $\boldsymbol{C}^{\ell,A}$ or $\boldsymbol{s}^{A}$) is also chiral.
For instance in bacteria such as \emph{E. coli}, the flagellar bundle tends to rotate the fluids anti-clockwise whereas the body tends to rotate the fluid clockwise.
Effectively, this can be seen as a `torque dipole' [see Fig.~\ref{fig:bacteria}(b)].
Indeed a `torque dipole' is the simplest realization of a \emph{chiral} active particle.

\subsection{There is no unique hydrodynamic description of active torque dipoles \label{CaseI}}
 
Microscopically, as we can see from Fig.~\ref{fig:bacteria}(c,d) (not an exhaustive list), there are multiple ways of representing a `torque dipole'.
For example, in the case of a bacterium, one can represent it as a composite of two equal and opposite point torques, separated by a distance $a$ (Case~Ia).
Alternatively, it can be described as a composite of four point forces (Case~Ib).
(Note that these point forces and torques are rigidly attached to a body frame set by $\vecr_i$ and $\hat{\vecp}_i$.)
In the limit of $b\rightarrow0$ and keeping $T=bF$ fixed, Case~Ia and Case~Ib might seem equivalent microscopically.
However, we will show that these two different representations of a torque dipole of given strength lead to different hydrodynamic descriptions and consequences. In effect one has to decide whether the torques involved act directly on the internal spin of particles (Case Ia), on CM degrees of freedom (Case Ib), or on a combination (Case II). 

First let us consider Case~Ia, as shown in in Fig.~\ref{fig:bacteria}(c).
The torque density [right hand side of (\ref{eq:elldot2})] arising from a suspension of torque dipoles is then
\begin{eqnarray}
\vecn\cdot\boldsymbol{C}^{\ell,A} &=& \left< \sum_i \left\{ T\hat{\vecp}_i\delta\left(\vecr - \vecr_i - \frac{a}{2}\hat{\vecp_i}\right)
									             -  T\hat{\vecp}_i\delta\left(\vecr - \vecr_i + \frac{a}{2}\hat{\vecp_i}\right) \right\} \right> \nonumber \\
						    &=& -Ta\vecn\cdot \left< \sum_i\hat{\vecp}_i\hat{\vecp}_i\delta(\vecr-\vecr_i) \right> 
						      = -Tan\vecn\cdot\left(\boldsymbol{Q}+\frac{\boldsymbol{I}}{d}\right),
\end{eqnarray}
where we have used the definition (\ref{eq:Q-micro}) of the nematic tensor $\boldsymbol{Q}$.
Therefore, in Case~Ia, we identify the active stress and active angular stress to be:
\begin{equation}
\sigma_{\alpha\beta}^A = 0 \quad\text{and}\quad C_{\alpha\beta}^{\ell,A} = -Tan\left[p_\alpha p_\beta+(1-|\vecp|^2)\frac{\delta_{\alpha\beta}}{3}\right], \label{eq:Ia}
\end{equation}
where we have used the approximation (\ref{eq:Q-approx}) to express $\boldsymbol{Q}$ in terms of $\vecp$.
Note that in this case the linear momentum equation (\ref{eq:udot2}) is not modified by the activity. Instead, 
the active term, $\boldsymbol{C}^{\ell,A}$, only affects the angular momentum equation (\ref{eq:elldot2}).

Next, we consider Case~Ib, as shown in in Fig.~\ref{fig:bacteria}(d).
We define two unit vectors, $\hat{\vecb}_i$ and $\hat{\vecc}_i$, such that
$\{\hat{\vecp}_i,\hat{\vecb}_i,\hat{\vecc}_i\}$
are orthonormal to each other, see Fig.~\ref{fig:bacteria}(d). 
Note that $\hat{\vecb}_i$ and $\hat{\vecc}_i$ are uniformly distributed on a plane perpendicular to $\vecp$ which must be true so long as the system remains uniaxial (as anyway required by the approximated form of $\boldsymbol{Q}$).
The force density arising from a suspension of these chiral force quadrupoles is:
\begin{align}
\vecn\cdot\boldsymbol{\sigma}^A &= \Bigg< \sum_i \bigg\{ F\hat{\vecc}_i\delta\left(\vecr - \vecr_i - \frac{a}{2}\hat{\vecp}_i - \frac{b}{2}\hat{\vecb}_i \right)
-  F\hat{\vecc}_i\delta\left(\vecr - \vecr_i - \frac{a}{2}\hat{\vecp}_i + \frac{b}{2}\hat{\vecb}_i \right)  \nonumber\\
&+ F\hat{\vecc}_i\delta\left(\vecr - \vecr_i + \frac{a}{2}\hat{\vecp}_i + \frac{b}{2}\hat{\vecb}_i \right)
-  F\hat{\vecc}_i\delta\left(\vecr - \vecr_i + \frac{a}{2}\hat{\vecp}_i - \frac{b}{2}\hat{\vecb}_i \right) \bigg\}\Bigg> \\
\Rightarrow\sigma_{\beta\gamma}^A &\simeq Fba\partial_\delta\left<\sum_i \hat{c}^i_{\beta}\hat{b}^i_{\gamma}\hat{p}^i_{\delta}\delta(\vecr-\vecr_i) \right> \, . 
\label{eq:sigmaA-1b-1}
\end{align}
Let us show that the symmetric part of $\boldsymbol{\sigma}^A$ is zero in this case.
Within the mesoscopic average there is no difference between $\hat\vecb_i$ and $\hat\vecc_i$ as both are uniformly distributed, hence, 
$\hat{\vecc}_{i}\hat{\vecc}_{i} = \half\left(\hat{\vecc}_{i}\hat{\vecc}_{i} + \hat{\vecb}_{i}\hat{\vecb}_{i}\right) = \half\left(\delta_{\alpha\beta} - \hat p^i_{\alpha}\hat p^i_{\beta}\right)$,
where the second equality is just the definition of the perpendicular projection of $\hat\vecp_i$ 
($\hat{\vecb}_{i}$ and $\hat{\vecc}_{i}$ span the plane perpendicular to $\hat\vecp_i$).
Then, using $\hat{\vecb}_i=\hat{\vecp}_i\times\hat{\vecc}_i$ and the above, 
we find that $\hat{c}^i_{\beta}\hat{b}^i_{\gamma}\hat{p}^i_{\delta} =  \varepsilon_{\gamma\beta\nu }\hat{p}^{i}_\delta \hat{p}^{i}_\nu$,
so that $\boldsymbol{\sigma}^A$ is  antisymmetric.
Multiplying (\ref{eq:sigmaA-1b-1}) with $\epsilon_{\alpha\beta\gamma}$ gives its anti-symmetric part as: 
\begin{equation}
\epsilon_{\alpha\beta\gamma} \sigma_{\beta\gamma}^{a,A} = -Ta \partial_\delta\left<\sum_i \hat{p}^i_{\alpha}\hat{p}^i_{\delta}\delta(\vecr-\vecr_i) \right>, \label{eq:sigmaA-1b}
\end{equation}
where we have used $T=Fb$ and $\hat{\vecp}_i=\hat{\vecb}_i\times\hat{\vecc}_i$.
Now multiplying (\ref{eq:sigmaA-1b}) again with $\epsilon_{\alpha\mu\nu}$, we identify the active stress $\sigma_{\alpha\beta}^A$
and the active angular stress $C_{\alpha\beta}^{\ell,A}$ for Case~Ib:
\begin{equation}
\sigma_{\alpha\beta}^{a,A} = \frac{1}{2}\epsilon_{\alpha\beta\gamma} 
				\partial_\pi\bigg\{\underbrace{-Tan\left[p_\gamma p_\pi + (1-|\vecp|^2)\frac{\delta_{\gamma\pi}}{3}\right]}_{C_{\gamma\pi}^{\sigma,A}}  \bigg\} \label{eq:Ib}
\quad\text{and}\quad C_{\alpha\beta}^{\ell,A}=0 \, .
\end{equation}
We end up in Case Ib with an anti-symmetric active stress that can be written purely as the divergence of another object.
Such an active stress only modifies the linear momentum equation (\ref{eq:udot2}) but not the angular momentum equation (\ref{eq:elldot2}),
since only anti-symmetric stress that cannot be written as a divergence, $\tilde{\boldsymbol{\sigma}}^{a,A}$, can enter into (\ref{eq:elldot2}) \emph{via} $\vecs^{A}$ (\ref{eq:sA}). The inability of the Case Ib torque dipole to affect the dynamics of $\vecl$ is in accord with the fact that it represents two force pairs, rather than two point torques.

Notice that $\C^{\ell,A}$ in (\ref{eq:Ia}) and $\C^{\sigma,A}$ in (\ref{eq:Ib}) have the same form.
Thus macroscopically, both Case~Ia and Case~Ib give the same surface torque, see equation (\ref{eq:Ldot2}) and Fig.~\ref{fig:surface-torque}.
However mesoscopically, we have two different hydrodynamic descriptions:
$\C^{\ell,A}$ only affects the angular momentum equation, 
whereas $\C^{\sigma,A}$ affects the linear momentum equation.
As we will show in Section~\ref{elimination}, after elimination of angular motion,
this will result in two different descriptions of the polar hydrodynamics.

\begin{figure}
\begin{centering}
\includegraphics[width=0.6\textwidth]{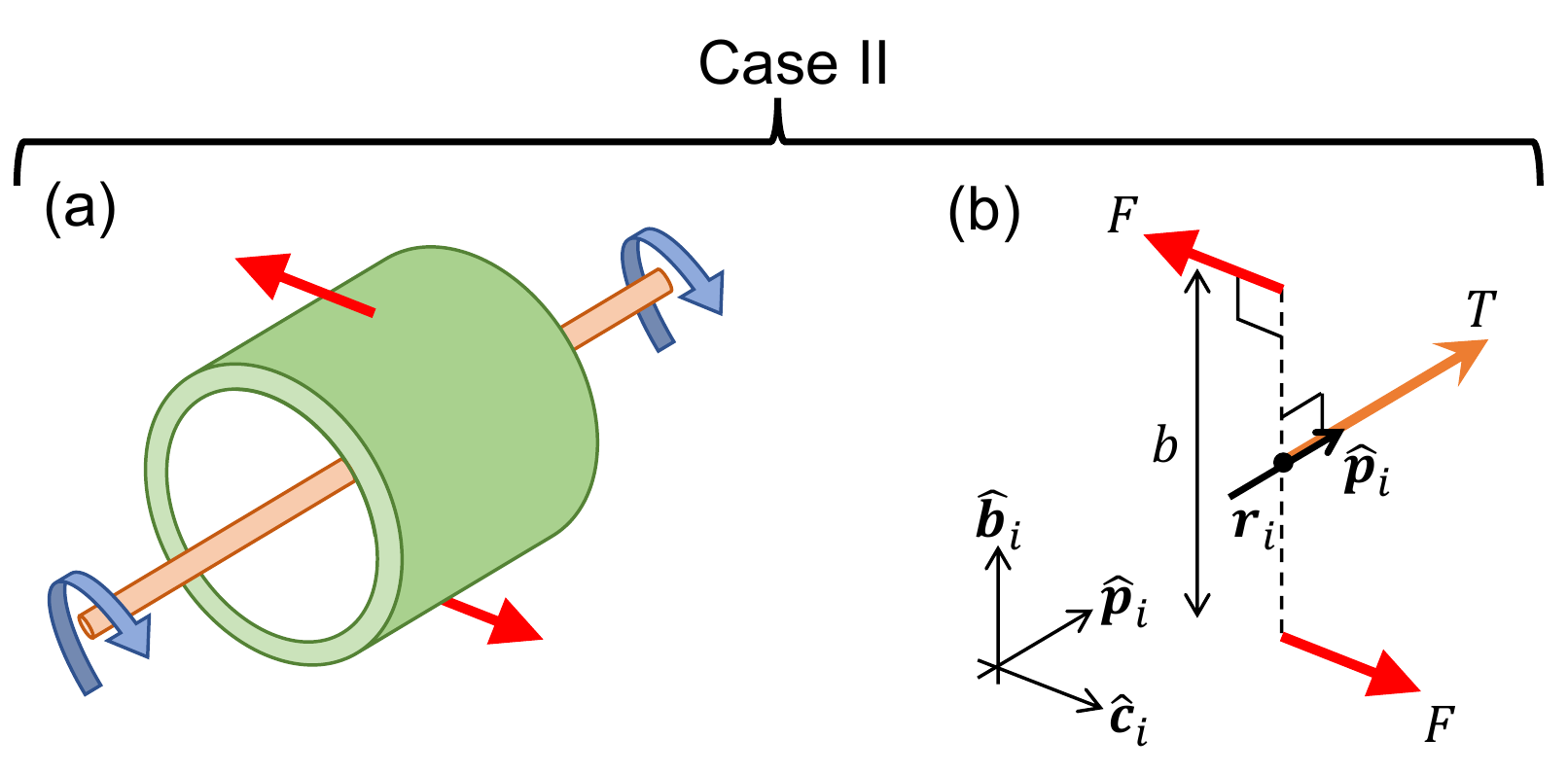}
\par\end{centering}
\caption{
(a) Another type of chiral active particle: counter-rotating cylinders (Case II).
(b) This particle can be approximated as a single torque monopole, balanced by a force pair. \label{fig:cylinder} }
\end{figure}

\subsection{Case II: An internal body torque and its physical interpretation\label{CaseII}}

Imagine that we now embed a point torque at $\vecr_i$ with magnitude $T$ in the direction of $\hat{\vecp}_i$ to each particle $i$ [Case~II as shown in Fig.~\ref{fig:cylinder}(b)]. 
To counter-balance this torque monopole, we add an equal and opposite force pair separated by a distance $b$, and centred at $\vecr_{i}$~\cite{furthauer}. 
The magnitude of this force is $F=T/b$. Thus, the net force and torque is zero as required.  
Physically, this chiral object might correspond to a rotating thin rod inside a counter-rotating cylinder, as shown in Fig.~\ref{fig:cylinder}(a). This could be a model of certain micro-organisms, such as spirochetes, with internal rather than external flagella.
As before, let us define two unit vectors $\hat{\vecb}_{i}$ and $\hat{\vecc}_{i}$ such that $\{\hat{\vecp}_{i},\hat{\vecb}_{i},\hat{\vecc}_{i}\}$ are orthonormal
[see Fig.~\ref{fig:cylinder}(b)]. 

First, the \emph{internal} torque density from a distribution of point torques at $\vecr_i$ is
\begin{eqnarray}
\vecs^A = \left<\sum_i T\hat{\vecp}_i \delta(\vecr-\vecr_i)\right>=Tn\vecp.
\end{eqnarray}
Next, the force density from the force pairs and hence the active stress is
\begin{eqnarray}
\vecn\cdot\boldsymbol{\sigma}^{A} &=& \left<\sum_i\left\{  F\hat{\vecc}_i\delta\left(\vecr - \vecr_i + \frac{b}{2}\hat{\vecb}_i\right)
									      -  F\hat{\vecc}_i\delta\left(\vecr - \vecr_i - \frac{b}{2}\hat{\vecb}_i\right)  \right\}\right> \\
\Rightarrow\sigma^{A}_{\alpha\beta} &=& T \left<\sum_i \hat{c}^i_{\alpha}  \hat{b}^i_{\beta}  \delta\left(\vecr - \vecr_i\right) \right>.
\end{eqnarray}
Multiplying with $\epsilon_{\alpha\beta\gamma}$, we get:
\begin{equation}
\epsilon_{\alpha\beta\gamma}\sigma^{A}_{\beta\gamma} = T \left<\sum_i \epsilon_{\alpha\beta\gamma}\hat{c}^i_{\beta}  \hat{b}^i_{\gamma}  \delta\left(\vecr - \vecr_i\right) \right> 
										       = -T \left<\sum_i \hat{p}^i_{\alpha}  \delta\left(\vecr - \vecr_i\right) \right> = -s_\alpha^A,
\end{equation}
since $\hat{\vecp}_i=\hat{\vecb}_i\times\hat{\vecc}_i$
(similar arguments as for Case~Ib shows that $\boldsymbol{\sigma}^{A}$ is antisymmetric). 
Thus we verify equality (\ref{eq:sA}) and hence the torque balance condition (\ref{eq:T-balance}).
Furthermore, we also show that in Case II the active stress arising from the force pairs is purely in a non-divergence form $\tilde{\boldsymbol{\sigma}}^A$.
This is because the active stress must exactly balance the internal body torque to ensure local cancellation of forces and torques for the active particle as a whole. 
Then, since the body torque has the stated form, so does the active stress.
Note that if the force pair is not centred at the position of the  point torque, $\vecr_i$, there is generally also contribution to $\C^{\ell,A}$ and $\C^{\sigma,A}$.
However, the constraint of local cancellation of torques~(\ref{eq:sA}) is always maintained.


\section{The limit of vanishing moment of inertia \label{elimination}}  

In this Section, we take the limit where the inertial timescale for the angular momentum is much smaller than the rotational viscous time scale. 
In this regime, we can take the limit of  $I\rightarrow0$ in order to eliminate the angular degree of freedom, $\boldsymbol{\Omega}$.
By doing so, we will derive an effective polar hydrodynamics for $\vecp(\vecr,t)$ and $\vecu(\vecr,t)$. We will do this first in general terms and then discuss the resulting differences between Cases Ia, Ib and II. 

We start from the full hydrodynamic equations, which we gather here as follows:
\begin{align}
\frac{Dp_{\alpha}}{Dt}+\Omega_{\alpha\beta}p_{\beta} &= - \gamma_{\alpha\beta}h_{\beta}
										    + \xi_1(\Omega_{\alpha\beta}-\omega_{\alpha\beta})p_\beta
										    + \xi_0\nu_{\alpha\beta}p_{\beta}
										    + \xi_2(\vecp\cdot\boldsymbol{\nu}\cdot\vecp)p_\alpha \label{eq:pdot3} \, ,\\
\rho\frac{Du_{\alpha}}{Dt} &= \partial_{\beta}\sigma_{\alpha\beta}^A+\partial_{\beta}\sigma_{\alpha\beta} \label{eq:udot3} \, , \\
\frac{D\ell_{\alpha}}{Dt} &= \partial_{\beta}C_{\alpha\beta}^{\ell,A} + s_\alpha^{A} - \epsilon_{\alpha\beta\gamma}\tilde{\sigma}_{\beta\gamma}^{a} \, . \label{eq:elldot3}
\end{align}
Here $\boldsymbol{\sigma}=\boldsymbol{\sigma}^{d}+\boldsymbol{\sigma}^{r}+\boldsymbol{\sigma}^{e}$
is the \emph{passive} stress, which consists of dissipative, reactive and elastic terms (\ref{eq:sigmad}-\ref{eq:sigmar});
$\tilde{\boldsymbol{\sigma}}^{a}$ is the part of $\boldsymbol{\sigma}$ that is antisymmetric, but not of divergence form, such that $-\epsilon_{\alpha\beta\gamma}\tilde{\sigma}_{\beta\gamma}^{a} = s_\alpha$ 
is the passive body torque. (Recall that there is no passive contribution to $\C^{\ell}$ in our model.) The active terms are
$\boldsymbol{\sigma}^{A}$, $\C^{\ell,A}$ and $\vecs^A$. 
For Case~Ia, of these only $\C^{\ell,A}$ is nonzero, while for Case~Ib the only non-vanishing active term is  $\sigma_{\alpha\beta}^{A}=\partial_{\pi}\left(\frac{1}{2}\epsilon_{\alpha\beta\gamma}C_{\gamma\pi}^{\sigma,A}\right)\neq0$. (Recall that in this case $\boldsymbol{\sigma}^{A}$ is purely of divergence form so that $\boldsymbol{s}^A = 0$ in (\ref{eq:elldot3}).) 
For Case~II the non-vanishing terms are $\boldsymbol{s}^A$ and $\boldsymbol{\sigma}^A$, which obey $s^A_{\alpha} = -\epsilon_{\alpha\beta\gamma}\tilde\sigma_{\beta\gamma}^{A} = -\epsilon_{\alpha\beta\gamma}\sigma_{\beta\gamma}^{A}$, with the second equality because $\boldsymbol{C}^{\sigma,A}$ now vanishes. 

To proceed, we assume for simplicity the rotational viscosity to be isotropic, 
$\eta'_{\alpha\beta\gamma\delta}=\frac{\eta'}{2}\epsilon_{\alpha\beta\mu}\epsilon_{\gamma\delta\mu}$.
Setting the left hand side of (\ref{eq:elldot3}) to zero we have 
\begin{equation}
\Omega_{\alpha\beta} = \omega_{\alpha\beta} 
				   - \frac{\eta_c}{2\eta'}(\nu_{\alpha\nu}p_\nu p_\beta - \nu_{\beta\nu}p_\nu p_\alpha)
				   - \frac{1-\xi_1}{2\eta'}(p_\alpha h_\beta - p_\beta h_\alpha) + \frac{1}{\eta'} {\cal T}^A_{\alpha\beta} \, ,
				   \label{eq:Omega2}
\end{equation}
where we define 
\begin{equation}{\cal T}^A_{\alpha\beta} \equiv \half\epsilon_{\alpha\beta\pi} \left( \partial_{\sigma}C_{\pi\sigma}^{\ell,A} + s^A_\pi \right)\label{eq:calT}\,.
\end{equation}
Substituting (\ref{eq:Omega2}-\ref{eq:calT}) into (\ref{eq:pdot3}-\ref{eq:udot3}), we get an effective polar hydrodynamics for $\vecp(\vecr,t)$ and $\vecu(\vecr,t)$
that reads
\begin{align}
\frac{Dp_{\alpha}}{Dt} + \omega_{\alpha\beta}p_{\beta} &= -\bar{\gamma}_{\alpha\beta}h_{\beta}
+ \bar{\xi}_{0}\nu_{\alpha\beta}p_{\beta}
+ \bar{\xi}_{2}(\vecp\cdot\boldsymbol{\nu}\cdot\vecp)p_{\alpha} 
- \frac{(1-\xi_{1})}{\eta'}{\cal T}^A_{\alpha\beta}p_{\beta} \, , \label{eq:pdot-effective}\\
\rho\frac{Du_{\alpha}}{Dt} &= \partial_{\beta}\bigg\{\bar{\eta}_{\alpha\beta\gamma\delta}\nu_{\gamma\delta}
+ \frac{\bar{\xi}_{0}}{2}(p_{\alpha}h_{\beta} + p_{\beta}h_{\alpha})
+ \frac{\bar{\xi}_{2}}{2}(\vecp\cdot\vech)p_{\alpha}p_{\beta}
- \frac{1}{2}(p_{\alpha}h_{\beta}-p_{\beta}h_{\alpha}) \nonumber\\
&+ \sigma_{\alpha\beta}^{e}
+ \sigma_{\alpha\beta}^{A} 
+ {\cal T}^A_{\alpha\beta} + \frac{\eta_{c}}{2\eta'}\left({\cal T}^A_{\alpha\gamma} p_{\gamma}p_{\beta} + {\cal T}^A_{\beta\gamma} p_{\gamma}p_{\alpha}\right) \bigg\} \, . \label{eq:udot-effective}
\end{align}

Dealing first with the passive terms, these correspond directly to the standard hydrodynamic description of polar particles \cite{elsen2018}, subject to the addition of the flow-stretch coupling first considered in \cite{shear}, but with the following renormalizations of the coefficients:
\begin{align}
\bar{\eta}_{\alpha\beta\gamma\delta} &= \eta_{\alpha\beta\gamma\delta}
							      - \frac{\eta_{c}^{2}}{4\eta'}(\delta_{\alpha\gamma}p_{\beta}p_{\delta}|\vecp|^{2}
							     + \delta_{\beta\gamma}p_{\alpha}p_{\delta}|\vecp|^{2} - 2p_{\alpha}p_{\beta}p_{\gamma}p_{\delta}) \label{eq:etabar} \, ,\\
\bar{\gamma}_{\alpha\beta} &= \gamma_{\alpha\beta}
					     + \frac{(1-\xi_1)^2|\vecp|^{2}}{2\eta'}\left(\delta_{\alpha\beta} - \frac{p_{\alpha}p_{\beta}}{|\vecp|^{2}}\right) \label{eq:gammabar} \, ,\\
\bar{\xi}_{0} &= \xi_{0}+\frac{\eta_{c}(1-\xi_{1})|\vecp|^{2}}{2\eta'} \label{eq:xi0bar} \, ,\\
\bar{\xi}_{2} &= \xi_{2}-\frac{\eta_{c}(1-\xi_{1})}{2\eta'}\, . \label{eq:xi2bar}
\end{align}
One can directly observe from these results that the dissipative/reactive nature of the different terms may change during the elimination process.
In the specific case considered here, $\eta_c$ was clearly related to dissipative terms in the general formalism; see (\ref{eq:sigmad}).
However, after elimination of angular momentum, it emerges that $\eta_c$ is related to a strictly reactive term in (\ref{eq:pdot-effective}),
but contributes to both reactive and dissipative terms in (\ref{eq:udot-effective}).
Note that the elastic stress, $\boldsymbol{\sigma}^e$, in (\ref{eq:udot-effective}) remains unchanged as it is defined to be the stress in response to purely elastic deformation 
(see~\ref{app-expansion}).

We now turn to the active terms, which enter via terms in $\boldsymbol{\sigma}^A$ and $\boldsymbol{{\cal T}}^A$, the latter as defined in \eqref{eq:calT}. Equation~(\ref{eq:pdot-effective})  first shows that $\boldsymbol{{\cal T}}^A$ gives an effective torque directly in the dynamics of $\vecp$.
Clearly, only the direction of $\vecp$ is affected, not its magnitude; however, we see that for $\xi_1 >1$ the active torque will rotate $\vecp$ in the ``wrong" way,
{\it i.e.}, the macroscopic handedness is opposite to the microscopic one. (See \cite{elsen2017} for a related phenomenon.) Notice that, when present, this terms should survive even in the limit of a dry system, whereby $\vecu(\vecr,t) \to 0$.
The presence or absence of the $\boldsymbol{\sigma}^A$ and $\boldsymbol{{\cal T}}^A$ terms depends however on the microscopic modelling of `torque dipoles', as we now describe. 

Consider first Case~Ib of Fig.~\ref{fig:bacteria}(d), where we represent each single bacterium as a chiral force quadrupole,  
In this case, $\boldsymbol{{\cal T}}^A=0$ in (\ref{eq:pdot-effective}-\ref{eq:udot-effective}). 
The only non-equilibrium term is the anti-symmetric active stress $\boldsymbol{\sigma}^A$ in (\ref{eq:udot-effective}), 
which remains unchanged under the elimination of angular momentum.
(The expression for $\boldsymbol{\sigma}^A$ is given in (\ref{eq:Ib}).)
This corresponds to the hydrodynamic theory of chiral active fluids considered in~\cite{elsen2017}.
Although the hydrodynamic theory in Ref.~\cite{elsen2017} was built phenomenologically, without consideration of internal angular momentum, we see now that it is full and proper description of Case Ib.
Accordingly, from~\cite{elsen2017}, we know that upon increasing the activity parameter (the magnitude of the torque $T=Fb$),
the homogenous solution $\vecp=$ constant and $\vecu=0$ becomes unstable.
In steady state, we get a spontaneous flow state where $\vecu\neq0$ and $\vecp$ acquires spontaneous twist deformation; we explore this further elsewhere~\cite{instability}. 

In contrast, for Case~Ia in Fig.~\ref{fig:bacteria}(c), where we represent each single bacterium as two point torques,
$\boldsymbol{\sigma}^A=0$ but $\C^{\ell,A}\neq0$ so that $\boldsymbol{{\cal T}}^A$ is nonzero.
(The expression for $\C^{\ell,A}$ is given in (\ref{eq:Ia}).)
As can be seen from (\ref{eq:udot-effective}), the active term $\C^{\ell,A}$, which was originally in the angular momentum equation, now
appears as an anti-symmetric active stress, $\boldsymbol{{\cal T}}^A$, in the Navier-Stokes equation (\ref{eq:udot-effective}) 
after elimination of the angular momentum.
This corresponds to the hydrodynamic theory of chiral active fluids considered in~\cite{furthauer}, 
except that Ref.~\cite{furthauer} did not fully eliminate the angular momentum equation to obtain the effective polar hydrodynamics as we are considering here.

Note that the active stress $\boldsymbol{{\cal T}}^A$ for Case~Ia in (\ref{eq:udot-effective}) has exactly the same form as $\boldsymbol{\sigma}^A$ for Case~Ib.
However, there are two key differences:
(i) in Case~Ia, $\boldsymbol{{\cal T}}^A$ also gives rise to a symmetric active stress in the Navier-Stokes equation [see last term in (\ref{eq:udot-effective})],
and (ii) $\boldsymbol{{\cal T}}^A$ also gives an effective torque directly on the $\vecp$-dynamics [last term in (\ref{eq:pdot-effective})].
This extra active term in the $\vecp$-dynamics gives a very different hydrodynamic instability compared to Case~Ib.
For instance above some activity threshold, the active term in (\ref{eq:pdot-effective}) can spontaneously twist the polarization field $\vecp$
without affecting the fluid velocity $\vecu$; we explore this further in~\cite{instability}.

Finally in Case II, $\boldsymbol{\sigma}^A$ and $\boldsymbol{{\cal T}}^A$ are both nonzero. This case was discussed alongside Case Ia (albeit without full elimination of angular dynamics) in 
\cite{furthauer}, who reported that it gives rise to a positive or negative yield stress in simple shear flow: that is, the shear stress is offset by an amount whose sign depends on that of the activity coefficients and does not vanish in the absence of macroscopic flow.

\section{Adding noise to the equations of motion \label{noise}}  

So far we have derived the deterministic dynamics (\ref{eq:pdot3}-\ref{eq:elldot3}) from linear irreversible thermodynamics
and then taken the limit of vanishing moment of inertia to obtain (\ref{eq:pdot-effective}-\ref{eq:udot-effective}).
To deal with thermal fluctuations, we need to add noises to these deterministic equations
such that in the passive limit the Boltzmann weight $\sim \exp(-F_0/\kbt)$ is recovered for the steady state. In general this requires the noises to be multiplicative and one must specify how to evaluate these multiplicative noises which generally requires specification of an It\^{o} or Stratanovich interpretation, or something in between \cite{lau}.
Each interpretation of the stochastic equation of motion is associated with a different Langevin equation whose corresponding Fokker-Planck equation recovers the Boltzmann distribution at steady state, 
so long as the correct noise is chosen~\cite{lau,basu}. In the process though, there generically arises a so-called `spurious drift' which is a deterministic term in the stochastic equation proportional to the noise variance (hence to $\kbt$). In many well known cases this spurious drift vanishes, but as shown below, this is not the case for the equations under discussion here, which we interpret in the It\^{o} sense throughout. 

Adding noise to the $\vecp$-dynamics  yields (see~\ref{app-drift} for details),
\begin{align}
\frac{D p_\alpha}{D t} + \omega_{\alpha\beta}p_\beta &= -\bar{\gamma}_{\alpha\beta}h_\beta  
									+ \bar{\xi}_0 \nu_{\alpha\beta}p_\beta
									+ \bar{\xi}_2 (\vecp\cdot\boldsymbol{\nu}\cdot\vecp) p_\alpha 
									- \frac{1-\xi_1}{\eta'}{\cal T}^A_{\alpha\beta}p_\beta \nonumber\\
									         &+ \Lambda_\alpha(\vecr,t) 
		+ k_{B}T\delta(\mathbf{0})\left[\bar{g}_{1}'(|\vecp|^{2})+\bar{g}_{2}'(|\vecp|^{2})|\vecp|^{2}+\bar{g}_{2}(|\vecp|^{2})\right]p_{\alpha} \, , \label{eq:pdot-noise}
\end{align}
where we have written 
$\bar{\gamma}_{\alpha\beta} = \bar{g}_1(|\vecp|^2) \delta_{\alpha\beta} + \bar{g}_2(|\vecp|^2) p_\alpha p_\beta$, 
as defined in (\ref{eq:gammabar}) and denote ${\cal O}'(x) = \D {\cal O}/ \D x$.
In the above equation we have introduced a thermodynamic Gaussian white noise, $\Lambda_\alpha$,
which has zero mean and variance that obeys detailed balance in the passive limit, $\left< \Lambda_\alpha\left(\vecr,t\right)\Lambda_\beta\left(\vecr',t'\right) \right> = 2\kbt\bar\gamma_{\alpha\beta}\delta(\vecr-\vecr')\delta(t-t')$.
Note that the spurious drift is proportional to $\delta(\boldsymbol{0}) = 1/a^3$, with $a$ being a short cutoff lengthscale. Even if it looks peculiar, this divergence is a well understood and generic property of the spurious drift in stochastic field theories; see \ref{app-drift} \cite{sancho,sancho2003}.

In the same way, we find that adding noise to the Navier-Stokes equation (\ref{eq:udot-effective}) gives,
\begin{align}
\rho\frac{Du_\alpha}{Dt} &= \partial_\beta \bigg\{ \bar{\eta}_{\alpha\beta\gamma\delta}\nu_{\gamma\delta}
					+   \frac{\bar{\xi}_0}{2}(p_\alpha h_\beta + p_\beta h_\alpha)
					+   \frac{\bar{\xi}_2}{2}(\vecp\cdot\vech)p_\alpha p_\beta
					-    \frac{1}{2}(p_\alpha h_\beta - p_\beta h_\alpha) 
					+   \sigma^e_{\alpha\beta} \nonumber\\
				      &+ \sigma^A_{\alpha\beta} 
				      +  {\cal T}^A_{\alpha\beta}
				      +   \frac{\eta_{c}}{2\eta'}\left({\cal T}^A_{\alpha\gamma}p_{\gamma}p_{\beta} + {\cal T}^A_{\beta\gamma}p_{\gamma}p_{\alpha} \right) 
				        +  \Sigma_{\alpha\beta}(\vecr,t)  \nonumber\\
				     &-  k_{B}T\delta(\mathbf{0})\left[ \bar{\xi}_{0}\delta_{\alpha\beta}
										     + (2\bar{\xi}_{0}' + 2\bar{\xi}_2'|\vecp|^2 + 3\bar{\xi}_2)p_{\alpha}p_{\beta} \right] \bigg\}, \label{eq:udot-noise}                   
\end{align}
with $\Sigma_{\alpha\beta}$ being a Gaussian noise with zero mean and variance
$\left< \Sigma_{\alpha\beta}(\vecr,t) \Sigma_{\gamma\delta}(\vecr',t') \right> = 2\kbt\bar{\eta}_{\alpha\beta\gamma\delta}\delta(\vecr-\vecr')\delta(t-t')$.
The last term in (\ref{eq:udot-noise}) is the spurious drift (see~\ref{app-drift}).
Importantly, the form of the the drift above is similar to the commonly used active stress in active liquid crystals, see~\cite{marchetti,elsen2015,hatwalne} 
(denoted here as $\boldsymbol{\sigma}^A$).
Although this spurious drift (with divergent prefactor) might then appear to dominate over any such active stress term, this is not actually the case since
the drift makes sure that the steady-state solution is Boltzmann distributed while the active stress (alongside any other active terms) makes sure that it isn't.
However, the presence of two terms of the same form does require care when simulating such systems: failing to consider the spurious drift 
might give completely wrong results.
Note that a similar form of spurious drift is present for any discretization scheme.

\section{Conclusion \label{conclusion}}                      

We have presented a rather general theory of active and passive polar liquid crystals with chiral activity, using linear irreversible thermodynamics and starting with an explicit treatment of spin angular momentum (which is later eliminated).

Among results for the passive limit,
our theory confirms the presence of the shear-elongation parameter and a cross-coupling viscosity.
The shear-elongation parameter was recently shown to play a role in shear-induced phase transitions in both passive and active cases~\cite{shear}.
The effects of the cross-coupling viscosity remain unstudied and merits further investigation. A further feature of the elimination of spin is the wholesale renormalization of the hydrodynamic parameters in the passive sector via equations (\ref{eq:etabar}--\ref{eq:xi2bar}). 

Our new results for the active case are relevant to many biological systems such as bacterial suspensions and actomyosin networks.
In many such cases the ultimate source of such activity is chiral.
The simplest implementation of this concept involves treating active particles as `torque dipoles'. However we have shown that there are at least three different hydrodynamic descriptions of polar liquid crystals subject to torque dipoles of a given strength, 
depending on whether the torque dipole are decomposed as two point torques (Case Ia), two force pairs (Case Ib), or as one of each (Case II); see Fig.~\ref{fig:bacteria}.

In most biological systems, the angular momentum can be regarded as a fast variable. 
Through proper elimination of angular degree of freedom,
we find, alongside the renormalization of parameters mentioned above, that the active stress is also modified in a non-trivial fashion. 
Specifically, chiral activity generically affects both the symmetric and antisymmetric parts of the stress tensor [see Eq.~(\ref{eq:udot-effective})].

Moreover, a direct effect of the chiral activity on the $\vecp$-dynamics (after elimination of angular momentum) is observed. 
When we treat the `torque dipole' as two force pairs, the $\vecp$-dynamics is not affected.
However, when the `torque dipole' is described as two point torques, we get an additional active chiral term in the $\vecp$-dynamics,
which tends to twist $\vecp$.
Consequently, these will give rise to different hydrodynamic instabilities, which will be investigated further in~\cite{instability}. The appearance of active torques directly in the dynamics of $\vecp$ opens various research directions, especially (but not exclusively) for `dry' active polar materials
such as bacteria on a rigid surface or in a gel matrix. 
We found that, depending on the sign of $\xi_1$ in the last term of Eq.~(\ref{eq:pdot-effective}),
we may get opposite macroscopic and microscopic handedness.
Although peculiar, such a behaviour is permitted by the Onsager symmetry and indeed was previously found in other systems~\cite{elsen2017,tombolato,vanRoij}.
Exploring the reversing of handedness through microscopic modelling is an appealing avenue for future study.

Meanwhile, the question of which of the two descriptions (Case Ia or Ib) is more appropriate for specific micro-organisms might be answered by seeing which instability is experimentally observed. They are not the only possibilities, with Case II (possibly relevant to endoflagellated microorganisms) being a third one. Further combinations, with (say) an unbalanced pair of torques balanced by a force pair, are easily envisaged and, within our formalism, their consequences are calculable. Likewise it is easy to add to either Case Ia or Ib a simple torque-free force dipole giving the traditional, symmetric form of active stress. This is likely necessary for bacteria and also other swimming microorganisms such as algae for which the existence of a stresslet flow field around the particle leave little doubt about the presence of a symmetric active stress \cite{drescher} without necessarily excluding the more complicated terms arising from chirality.

It would be of interest to investigate further the active chiral terms found in this work and 
study in more detail their effects on the dynamics and stability of both `dry' and `wet' active fluids, which are relevant for numerous biological systems. A further role of chirality can enter via the static free energy leading to the emergence of cholesteric rather than polar order \cite{chaikin}. A study of the interaction between structural and active chirality is likely to be complicated, but suggests further new avenues for research.

{\it Acknowledgements.~}
We thank E. Fodor, T.~C. Lubensky and R. Mari for useful discussions.
TM acknowledges support from the Blavatnik postdoctoral fellowship programme
and the National Science Foundation Center for Theoretical Biological Physics (Grant PHY-1427654). Work funded in part by the European Research Council under the Horizon 2020 Programme, ERC grant agreement number 740269. MEC is funded by the Royal Society.

\appendix

\section{General moment of inertia and compressible fluid \label{app-inertia}}

In this Appendix we describe the changes required to address both a general moment of inertia replacing \eqref{eq:Omega}, and the case of a compressible fluid.
Assuming the moment of inertia density to depend only on coarse grained variables (density and polarization) we can write it generally in the form
\begin{eqnarray}
I_{\alpha\beta} = n \left[\Delta I(|\vecp|^2) \frac{p_\alpha p_\beta}{|\vecp|^2} + I_\perp(|\vecp|^2)\delta_{\alpha\beta}\right] \, , \label{eqB:I}
\end{eqnarray}
where $I_\parallel$ and $I_\perp$ are the average moment of inertia in the directions parallel and perpendicular to $\vecp$, respectively.
The difference $\Delta I = I_\parallel - I_\perp$ characterizes the anisotropy of the average moment of inertia.
Note that in a microscopic calculation,  $I_{\alpha\beta}$ will depend on
the nematic order parameter $\boldsymbol{Q}$~\cite{lubensky2005}.
So long as $\boldsymbol{Q}$ can itself be written as a function of $\vecp$ only (for example by using the approximation of equation~\ref{eq:Q-approx}) the form \eqref{eqB:I} must result. 

Allowing for compressibility, the continuity equation for the mass density $\rho$ replaces the fluid incompressibility condition $\vecn\cdot\vecu = 0$:
\begin{equation}
\frac{\partial\rho}{\partial t} + \vecn\cdot\left(\rho\vecu\right) = 0\,.
\end{equation}
The free-energy (\ref{eq:F}) is also modified to include an equation of state and square gradient theory with respect to the density variable,
\begin{eqnarray}
\label{I2}
F[\rho,\vecp,\vecu,\vecl] = \int \D V \left[ \frac{1}{2}\rho u^2 
+ \frac{1}{2} \ell_\alpha I^{-1}_{\alpha\beta} \ell_{\beta}
+ \mathbb{F}_0 \left(\rho,\vecn\rho,\vecp,\vecn\vecp\right)  \right] \, .
\end{eqnarray}
The elastic stress now becomes:
\begin{equation}
\sigma_{\alpha\beta}^e = (\mathbb{F}_0 - \rho\mu - \vecp\cdot\vech)\delta_{\alpha\beta}
- \frac{\partial \mathbb{F}_0}{\partial(\partial_\beta\rho)}(\partial_\alpha\rho)
- \frac{\partial \mathbb{F}_0}{\partial(\partial_\beta p_\gamma)}(\partial_\alpha p_\gamma),
\end{equation}
where $\mu=\frac{\delta F_0}{\delta\rho}$ is the chemical potential (with $F_0$ as usual the configurational part of $F$)
and the corresponding Gibbs-Duhem relation becomes
\begin{equation}
\partial_\beta\sigma^e_{\alpha\beta} = - \rho\partial_\alpha\mu - p_\beta\partial_\alpha h_\beta.
\end{equation}
The antisymmetric part of the elastic stress tensor is now modified by $\Delta I$ and reads,
\begin{eqnarray}
\sigma_{\alpha\beta}^{a,e} &=& \frac{1}{2}(p_\alpha h_\beta - p_\beta h_\alpha) 
- \frac{1}{2} (\Omega_\alpha\ell_\beta - \Omega_\beta\ell_\alpha)
- \frac{\Delta I}{2|\vecp|^2} \left(\vecp \cdot \boldsymbol{\Omega}\right)(p_\alpha\Omega_\beta - p_\beta\Omega_\alpha) \nonumber\\
&+& \partial_\gamma\left[ \frac{1}{2}\left(\frac{\partial \mathbb{F}_0}{\partial(\partial_\gamma p_\beta)}p_\alpha  
- \frac{\partial \mathbb{F}_0}{\partial(\partial_\gamma p_\alpha)}p_\beta \right) \right] \, .
\end{eqnarray}
The equations of motion derived in \ref{app-expansion} above are all unchanged except for the replacement
\begin{equation}
h_\alpha \rightarrow h_\alpha - \frac{1}{2}\Omega_\gamma\frac{\partial I_{\gamma\delta}}{\partial p_\alpha}\Omega_\delta
\end{equation}
in all kinetic coefficients.
Finally, the reactive terms related to $\chi_1^\nu$ in (\ref{eqA:Xomega}) are no longer zero.
In deriving the above equation we have used the form (\ref{eqB:I}) for the moment of inertia,
and the resulting expression for the derivative of its inverse,
$\partial_x I_{\alpha\beta}^{-1} = I_{\alpha\gamma}^{-1} (\partial_x I_{\gamma\delta}) I_{\delta\beta}^{-1} $.

\section{Systematic expansion of Onsager coefficients \label{app-expansion}}

From the conservation laws (\ref{eq:udot}) and (\ref{eq:elldot}) we have, for equilibrium,
\begin{align}
\rho\frac{Du_{\alpha}}{Dt} & = \partial_\beta\sigma_{\alpha\beta}^{e}
					   + \partial_\beta\sigma_{\alpha\beta}^{s,k}
					   + \partial_\beta\sigma_{\alpha\beta}^{a,k}
					   + f_{\alpha} \label{eqA:udot} \, , \\
\frac{D\ell_{\alpha}}{Dt} & = -\epsilon_{\alpha\beta\gamma}\tilde\sigma_{\beta\gamma}^{a,e}
				        - \epsilon_{\alpha\beta\gamma}\tilde\sigma_{\beta\gamma}^{a,k}
				        + \partial_\beta C_{\alpha\beta}^{\ell} 
				        + \tau_\alpha\, . \label{eqA:ldot}
\end{align}
We have split the stress tensor $\boldsymbol{\sigma}$ into elastic parts (superscript `$e$') and a kinetic part (dissipative and reactive -- superscript `$k$'). 
Superscripts `$s$' and `$a$' indicate symmetric and anti-symmetric parts for the particular $\boldsymbol{\sigma}$.
The tildes refer to parts of the anti-symmetric stresses that correspond to body torques and cannot be written as the divergence of a third-rank tensor.
Note that only these non-divergence parts of $\boldsymbol{\sigma}$ appear in the angular momentum equation (\ref{eqA:ldot}). 
The elastic and reactive terms do not contribute to entropy production, only the dissipative terms do. 
Note that $\C^\ell$ does not have an elastic part but only dissipative and reactive contributions. 

Now the dynamics of $\vecp$ can be written as:
\begin{equation}
\frac{\partial p_{\alpha}}{\partial t}+\left(\boldsymbol{u}\cdot\vecn\right)p_{\alpha}+\Omega_{\alpha\beta}p_{\beta} = \ldots. \label{eqA:pdot}
\end{equation}
The left hand side in the equation above follows from Galilean and rotational invariance, and allows for advection with the fluid of the CM of each particle and rotation of its orientation $\hat\vecp_i$ caused by its spin $\vecl_i$, with $\Omega_{\alpha\beta}$ defined such that $\Omega_{\alpha\beta}p_\beta$ is the resulting rotational velocity of the coarse grained polarization $\vecp(\vecr,t)$. The rotation rate tensor $-\Omega_{\alpha\beta}$ is as usual related via $\Omega_{\alpha\beta} = \epsilon_{\alpha\beta\gamma} \Omega_\gamma$ 
to the angular velocity vector $\Omega_\gamma$ which we take to obey \eqref{eq:Omega}.

In this Appendix, we will derive the right hand side of (\ref{eqA:pdot}), 
together with $\boldsymbol{\sigma}^{e}$, $\boldsymbol{\sigma}^{k}$, and $\C^{\ell}$ systematically, 
taking into account the Onsager reciprocal relations.
In particular, we will derive the general form of the Onsager coefficients including all chiral terms, while disregarding higher order gradient contributions.

The first step is to derive the elastic or Ericksen stress $\boldsymbol{\sigma}^{e}$ in the Navier-Stokes equation (\ref{eqA:udot}). 
Suppose we strain the material by an arbitrarily small amount $\delta\vecr$: $\vecr\rightarrow\vecr+\delta\vecr$.
The volume of a patch of the material will also change as a result: $V\rightarrow V+\delta V$.
The change in the configurational free energy is:
\begin{align}
\nonumber\delta F_{0} & = \int_{V+\delta V} \mathbb{F}_0(\vecp+\delta\vecp,\vecn\vecp+\vecn\delta\vecp) \, \D V
		      - \int_V \mathbb{F}_0(\vecp,\vecn\vecp) \, \D V \\
\nonumber		  & = \int_V \Big[ \mathbb{F}_0(\vecp+\delta\vecp,\vecn\vecp+\vecn\delta\vecp) - \mathbb{F}_0(\vecp,\vecn\vecp) \Big] \D V
		     + \oint_{\partial V} \left[ \mathbb{F}_0\delta r_\beta \right] \D S_\beta \\
 		  & = \int_V \left[ h_\alpha\delta p_\alpha \right] \D V
		     + \oint_{\partial V} \left[ \mathbb{F}_0\delta r_\beta + \frac{\partial \mathbb{F}_0}{\partial(\partial_\beta p_\alpha)}\delta p_{\alpha} \right] \D S_\beta, \label{eqA:deltaF1}
\end{align}
where $\boldsymbol{h}=\frac{\delta F_{0}}{\delta\vecp}$ is the molecular field and $\mathbb{F}_0$ is the configurational free energy density.
For \emph{affine} deformation, the change in the polarization is given by $\delta\vecp=-(\delta\vecr\cdot\vecn)\vecp.$
Then (\ref{eqA:deltaF1}) becomes:
\begin{equation}
\delta F_{0} = \int_V \left[ p_{\beta}\partial_{\alpha}h_{\beta} \right] \delta r_{\alpha}\, \D V
		  + \oint_{\partial V} \delta r_\alpha \left[ (\mathbb{F}_0-\vecp\cdot\boldsymbol{h})\delta_{\alpha\beta} - \frac{\partial \mathbb{F}_0}{\partial(\partial_\beta p_\gamma)}\partial_\alpha p_\gamma \right] \D S_\beta. \label{eqA:deltaF2}
\end{equation}

On the other hand, the work done on the system is the product of elastic stress and strain (see~\cite{LL}):
\begin{equation}
\delta W = \int_{V} \sigma_{\alpha\beta}^e \frac{\partial\delta r_\alpha}{\partial r_\beta} \, \D V 
	      = -\int_{V}\left( \partial_\beta\sigma_{\alpha\beta}^e \right)\delta r_\alpha \, \D V
	      + \oint_{\partial V} \delta r_\alpha \sigma_{\alpha\beta}^e \,\D S_\beta .\label{eqA:deltaW}
\end{equation}
In the first equality, $\partial\delta r_{\alpha}/\partial r_\beta \equiv \partial_{\beta}\delta r_{\alpha}$ is simply the strain tensor. 
Since the change is reversible, $\delta F_{0}=\delta W$ and thus equating the surface term of (\ref{eqA:deltaW}) to that of (\ref{eqA:deltaF2}), 
we get the elastic stress
\begin{equation}
\sigma_{\alpha\beta}^e = \left(\mathbb{F}_0 - \vecp\cdot\boldsymbol{h}\right) \delta_{\alpha\beta}
				    - \frac{\partial \mathbb{F}_0}{\partial(\partial_\beta p_\gamma)}\left(\partial_\alpha p_\gamma\right)\,. \label{eqA:sigmae}
\end{equation}
while equating the volume terms gives the Gibbs-Duhem relation
\begin{equation}
\partial_\beta\sigma_{\alpha\beta}^e = -p_\beta\partial_\alpha h_\beta.\label{eqA:divsigmae}
\end{equation}
What remains is to find the anti-symmetric part of the elastic stress, $\boldsymbol{\sigma}^{a,e}$. 
To derive this, suppose we rotate the whole material as a rigid body with some small angle $\boldsymbol{\theta}$: $\vecr\rightarrow\boldsymbol{\theta}\times\vecr$.
The displacement field is $\delta\vecr=\boldsymbol{\theta}\times\vecr$ and the change in the polarization field is given by (for rigid body rotation)
\begin{equation}
\delta\vecp = -\delta\vecr\cdot\vecn\vecp + \boldsymbol{\theta}\times\vecp. \label{eqA:deltap}
\end{equation}
Substituting (\ref{eqA:deltap}) to (\ref{eqA:deltaF1}) and using (\ref{eqA:sigmae}-\ref{eqA:divsigmae}) we get
\begin{align}
\nonumber\delta F_{0} & = \int_{V}\left[h_{\alpha}\epsilon_{\alpha\pi\beta}\theta_{\pi}p_{\beta}
		      + \sigma_{\alpha\beta}^{e}\frac{\partial\delta r_{\alpha}}{\partial r_{\beta}}\right] \D V
		      + \oint_{\partial V}\left[\frac{\partial \mathbb{F}_{0}}{\partial(\partial_{\gamma}p_{\alpha})}\epsilon_{\alpha\pi\beta}\theta_{\pi}p_{\beta}\right] \D S_\gamma \\
		  & = \int_{V}\Big[ -\frac{1}{2} \epsilon_{\alpha\beta\pi} \theta_\pi (h_\alpha p_\beta - h_\beta p_\alpha)
		  			 - \frac{1}{2} \epsilon_{\alpha\beta\pi}\theta_{\pi}\partial_{\gamma}\left(\frac{\partial \mathbb{F}_{0}}{\partial(\partial_{\gamma}p_{\alpha})}p_{\beta}-\frac{\partial \mathbb{F}_{0}}{\partial(\partial_{\gamma}p_{\beta})}p_{\alpha}\right) \nonumber \\
		  & - \sigma_{\alpha\beta}^{a,e}\epsilon_{\alpha\beta\pi}\theta_{\pi} \Big] \D V\label{eqA:deltaF3}
\end{align}
However, rigid body rotation should not change the free energy. 
Setting $\delta F_{0}=0$, one finds the anti-symmetric part of the elastic stress,
\begin{equation}
\sigma_{\alpha\beta}^{a,e} = \underbrace{\frac{1}{2}\left(p_{\alpha}h_{\beta}-p_{\beta}h_{\alpha}\right)}_{\tilde{\sigma}_{\alpha\beta}^{a,e}}
					 + \partial_\gamma \Bigg[ \frac{1}{2} 
					 \underbrace{\left( \frac{\partial \mathbb{F}_{0}}{\partial(\partial_{\gamma}p_{\beta})}p_{\alpha} 
					 			  - \frac{\partial \mathbb{F}_{0}}{\partial(\partial_{\gamma}p_{\alpha})}p_{\beta} \right)}_{\epsilon_{\alpha\beta\delta}C_{\delta\gamma}^{\sigma,e}}
				          \Bigg]. \label{eqA:sigmaae}
\end{equation}
Note that $\sigma_{\alpha\beta}^{a,e}$ is automatically decomposed
into a divergence part $\partial_\gamma\left[\frac{1}{2}\epsilon_{\alpha\beta\delta}C_{\delta\gamma}^{\sigma,e}\right]$ and a remainder $\tilde\sigma_{\alpha\beta}^{a,e}$, 
see (\ref{eq:sigmaa}). As discussed after \eqref{eq:sigmaae} above, we associate the first of these with an elastic body torque.

To complete the the equations of motion it is necessary to find the dissipative and reactive terms in (\ref{eqA:udot}-\ref{eqA:pdot}).
This can be done {\it via} the rate of entropy production. 
Suppose the isothermal polar fluid is in contact with a heat reservoir of temperature $T$. 
Let $S$ be the entropy of the system and $S_{r}$ be the entropy of the reservoir. 
The total entropy is then $S_{\rm tot}=S+S_{r}$. 
Now the free energy can be written as $F=E-TS$, where $E$ is internal energy.
The change in the system's free energy is 
\begin{equation}
\Delta F =\Delta E-T\Delta S=\Delta W+\Delta Q-T\Delta S,
\end{equation}
where $\Delta W$ is the work done on the system by the external forces and torques, 
$\Delta Q$ is the heat flux going into the system and 
$\Delta S$ is the increase in the system's entropy. 
Note that $\Delta Q\neq T\Delta S$, since we have friction in the system. 
The heat dissipated from the system into the reservoir, $-\Delta Q$, increases the entropy of the latter by $\Delta S_{r} = -\Delta Q/T$ (the reservoir itself is frictionless).
Therefore we find
\begin{equation}
\Delta F=\Delta W-T\Delta S_{r}-T\Delta S=\Delta W-T\Delta S_{\rm tot}.
\end{equation}
In other words, the total rate of entropy production (system + reservoir) is
\begin{equation}
T\frac{\D S_{\rm tot}}{\D t} = \frac{\D W}{\D t}-\frac{\D F}{\D t} 
			      = \int\left(\boldsymbol{f}\cdot\boldsymbol{u} + \boldsymbol{\tau}\cdot\boldsymbol{\Omega}\right) \D V - \frac{\D F}{\D t} \, . \label{eqA:Sdot1}
\end{equation}
The rate of change of free energy, including both configurational and kinetic parts, is
\begin{align}
\frac{\D F}{\D t} & =\int\left(\rho u_{\alpha}\frac{\partial u_{\alpha}}{\partial t}+\Omega_{\alpha}\frac{\partial\ell_{\alpha}}{\partial t}
+h_{\alpha}\frac{\partial p_{\alpha}}{\partial t}\right) \D V \, ,
\end{align}
and by substituting (\ref{eqA:udot}-\ref{eqA:ldot}), we get
\begin{align}
\frac{\D F}{\D t} & = \int\bigg[ u_{\alpha}\left(-u_{\beta}\partial_{\beta}u_{\alpha}+\partial_{\beta}\sigma_{\alpha\beta}^{e}
+\partial_{\beta}\sigma_{\alpha\beta}^{s,k}+\partial_{\beta}\sigma_{\alpha\beta}^{a,k}\right) \nonumber \\
 		   & + \Omega_{\alpha}\left(-u_{\beta}\partial_{\beta}\ell_{\alpha}-\epsilon_{\alpha\beta\gamma}\tilde{\sigma}_{\beta\gamma}^{a,e}
 		   -\epsilon_{\alpha\beta\gamma}\tilde{\sigma}_{\beta\gamma}^{a,k}+\partial_{\beta}C_{\alpha\beta}^{\ell}\right)\nonumber \\
 		   & + h_{\alpha}\left(\frac{Dp_{\alpha}}{Dt}+\Omega_{\alpha\beta}p_{\beta}\right)-h_{\alpha}u_{\beta}\partial_{\beta}p_{\alpha}
 		   -\Omega_{\alpha\beta}h_{\alpha}p_{\beta}+\boldsymbol{f}\cdot\boldsymbol{u}+\boldsymbol{\tau}\cdot\boldsymbol{\Omega}\bigg] \D V. \label{eqA:Fdot}
\end{align}
Using the Gibbs-Duhem relation (\ref{eqA:divsigmae}) and the formula for $\tilde{\boldsymbol{\sigma}}^{a,e}$ in (\ref{eqA:sigmaae}), 
we can eliminate all the elastic stresses in (\ref{eqA:Fdot}). 
Substituting (\ref{eqA:Fdot}) to (\ref{eqA:Sdot1}), and using integration by parts and the incompressibility condition $\vecn\cdot\boldsymbol{u}=0$, 
we obtain the total rate of entropy production:
\begin{eqnarray}
T\frac{dS_{tot}}{dt} &=& \int\Bigg[ -\left(\frac{Dp_{\alpha}}{Dt}+\Omega_{\alpha\beta}p_{\beta}\right)h_{\alpha}
			     			+ \nu_{\alpha\beta} \, \sigma_{\alpha\beta}^{s,k}
			     			+ \left(\Omega_{\alpha\beta}-\omega_{\alpha\beta}\right)\sigma_{\alpha\beta}^{a,k} \nonumber\\
			     &+& \left(C_{\alpha\beta}^{\ell} + C_{\alpha\beta}^{\sigma,k}\right)\partial_\beta\Omega_\alpha \Bigg] \D V \, . \label{eqA:Sdot2}
\end{eqnarray}
For the case of a compressible fluid, see \ref{app-inertia}.

Following established principles \cite{mazur}, the entropy production rate is next written as a sum of thermodynamic `fluxes' times `forces':
\begin{center}
\begin{tabular}{ccc}
Flux $\mathcal{J}$ &$\longleftrightarrow$& Force $\mathcal{F}$\tabularnewline 
\hline 
$\frac{Dp_{\alpha}}{Dt}+\Omega_{\alpha\beta}p_{\beta}$ ($-$) &$\longleftrightarrow$& $-h_{\alpha}$ ($+$)\tabularnewline
$\sigma_{\alpha\beta}^{s,k}$ ($+$) 					       &$\longleftrightarrow$& $\nu_{\alpha\beta}$ ($-$)\tabularnewline
$\sigma_{\alpha\beta}^{a,k}$ ($+$) 			       &$\longleftrightarrow$& $\Omega_{\alpha\beta}-\omega_{\alpha\beta}$ ($-$)\tabularnewline
$C_{\alpha\beta}^{\ell}+C_{\alpha\beta}^{\sigma,k}$ ($+$)  &$\longleftrightarrow$& $\partial_{\beta}\Omega_{\alpha}$ ($-$)\tabularnewline
\end{tabular}
\par\end{center}
Here the ($\pm$) sign indicates whether the quantity is even ($+$) or odd ($-$) under time reversal $t\rightarrow-t$. 
Within the framework of linear irreversible thermodynamics, the fluxes $\mathcal{J}_{i}$ are expanded to linear order in the forces $\mathcal{F}_{i}$,
while respecting Onsager reciprocal relations, yielding
\begin{align}
&\underbrace{\left( \begin{array}{c}
			    \frac{Dp_{\alpha}}{Dt}+\Omega_{\alpha\beta}p_{\beta} \\
			    \sigma_{\alpha\beta}^{s,k} \\
			    \sigma_{\alpha\beta}^{a,k} \\
			    C_{\alpha\beta}^{\ell} + C_{\alpha\beta}^{\sigma,k}
			    \end{array} \right)}_{\mathcal{J}_i} 
	 = \underbrace{\left(\begin{array}{cccc}
				\gamma_{(\alpha\pi)} &                     0 			              &                            0                                &                        0 \\
				               0		 & \eta_{(\alpha\beta)(\pi\sigma)}             & Y_{(\alpha\beta)[\pi\sigma]}^{\omega} & Y_{(\alpha\beta)\pi\sigma}^{\Omega} \\
					       0 		 & Y_{(\pi\sigma)[\alpha\beta]}^{\omega} & \eta'_{[\alpha\beta][\pi\sigma]}             &  Z_{[\alpha\beta]\pi\sigma} \\
				               0 		 & Y_{(\pi\sigma)\alpha\beta}^{\Omega}  & Z_{[\pi\sigma]\alpha\beta}                    & \zeta_{\alpha\beta\pi\sigma}
				\end{array}\right)}_{\mathcal{D}_{ij}}
	    \underbrace{\left(\begin{array}{c}
				-h_{\pi} \\
				\nu_{\pi\sigma}\\
				\Omega_{\pi\sigma}-\omega_{\pi\sigma}\\
				\partial_{\sigma}\Omega_{\pi}
				\end{array}\right)}_{\mathcal{F}_{j}} \nonumber \\
      & + \underbrace{\left(\begin{array}{cccc}
				                      0                         & X_{\alpha(\pi\sigma)}^{\nu} & X_{\alpha[\pi\sigma]}^{\omega} & X_{\alpha\pi\sigma}^{\Omega} \\
				-X_{\pi(\alpha\beta)}^{\nu}        &                       0                    &                            0                     &                    0  \\
				-X_{\pi[\alpha\beta]}^{\omega}  &                      0                    &                            0                     &                     0 \\
				-X_{\pi\alpha\beta}^{\Omega}   &                      0                     &                            0                     &                    0
				\end{array}\right)}_{\mathcal{R}_{ij}}
	    \underbrace{\left(\begin{array}{c}
				-h_{\pi} \\
				\nu_{\pi\sigma} \\
				\Omega_{\pi\sigma}-\omega_{\pi\sigma} \\
				\partial_{\pi}\Omega_{\sigma}
				\end{array}\right)}_{\mathcal{F}_{j}}. \label{eqA:flux-force}
\end{align}
Here round and square brackets indicate symmetric and anti-symmetric pair of indices respectively;
$\boldsymbol{\mathcal{D}}$ and $\boldsymbol{\mathcal{R}}$ are dissipative and reactive Onsager coefficients respectively; and 
$\boldsymbol{\mathcal{D}}$ is symmetric and positive semi-definite whereas $\boldsymbol{\mathcal{R}}$ is anti-symmetric. 
Only $\boldsymbol{\mathcal{D}}$ contributes to entropy production, since (\ref{eqA:Sdot2}) can now be written as
\begin{equation}
T\frac{\D S_{tot}}{\D t} = \int\mathcal{J}_{i}\mathcal{F}_{i}\,\D V
			      = \int\left[\mathcal{F}_{i}\mathcal{D}_{ij}\mathcal{F}_{j}+\mathcal{F}_{i}\mathcal{R}_{ij}\mathcal{F}_{j}\right] \D V
			      = \int\mathcal{F}_{i}\mathcal{D}_{ij}\mathcal{F}_{j}\,\D V \ge 0,
\end{equation}
which is positive semi-definite as expected. 
Notice that the reactive coefficients $\boldsymbol{\mathcal{R}}$ couples two thermodynamic quantities of the same time signature and this must not give rise to entropy production.
This can be proven from the equivalent Langevin dynamics (see~\cite{EPR,basu}). 
In \eqref{eqA:flux-force}, the quantities $\eta_{(\alpha\beta)(\gamma\delta)}$ and $\eta'_{[\alpha\beta][\gamma\delta]}$, which reside in $\boldsymbol{\mathcal{D}}$, are the shear and rotational viscosity respectively. 

In general, all coefficients inside $\boldsymbol{\mathcal{D}}$ and $\boldsymbol{\mathcal{R}}$ depend on $\vecp$. 
The forms for the diagonal viscosities in $\boldsymbol{\mathcal{D}}$ are found, in accord with Curie's symmetry principle~\cite{mazur}, as:
\begin{align}
\gamma_{(\alpha\beta)} & = g_{1}\delta_{\alpha\beta}+g_{2}p_{\alpha}p_{\beta}\label{eqA:gamma} \, ,\\
\eta_{(\alpha\beta)(\gamma\delta)}(\{\alpha_{i}\}) & = \alpha_{1}p_{\alpha}p_{\beta}p_{\gamma}p_{\delta}
									     + \alpha_{2}(\delta_{\alpha\gamma}\delta_{\beta\delta}+\delta_{\alpha\delta}\delta_{\beta\gamma}) \nonumber  \\
									  & + \alpha_{3}(p_{\alpha}p_{\gamma}\delta_{\beta\delta}+p_{\beta}p_{\gamma}\delta_{\alpha\delta}
									  +p_{\alpha}p_{\delta}\delta_{\beta\gamma}+p_{\beta}p_{\delta}\delta_{\alpha\gamma}) \nonumber \\
 									  & + \alpha_{4}\delta_{\alpha\beta}\delta_{\gamma\delta}
									     + \alpha_{5}(\delta_{\alpha\beta}p_{\gamma}p_{\delta}+\delta_{\gamma\delta}p_{\alpha}p_{\beta}) \label{eqA:eta} \, ,\\
\eta'_{[\alpha\beta][\gamma\delta]}(\{\alpha_{i}\}) & = \epsilon_{\alpha\beta\mu}\epsilon_{\gamma\delta\nu}\left(\alpha_{1}\delta_{\mu\nu}+\alpha_{2}p_{\mu}p_{\nu}\right) \, ,\\
\zeta_{\alpha\beta\gamma\delta}(\{\alpha_{i},\alpha_{i}'\}) & = \eta_{(\alpha\beta)(\gamma\delta)}(\{\alpha_{i}\})+\eta'_{[\alpha\beta][\gamma\delta]}(\{\alpha_{i}'\}) \, , \label{eqA:zeta}
\end{align}
where all the prefactors $\{g_i,\alpha_i,\alpha_i'\}$ are functions of $|\vecp|^2$ and $n$. 
Notice that $\gamma_{(\alpha\beta)}$ must be symmetric with respect to exchanging $\alpha\leftrightarrow\beta$,
thus, (\ref{eqA:gamma}) is the only possible expansion of $\boldsymbol{\gamma}$ in terms of $\vecp$. 
Similarly, the linear viscosity $\eta_{(\alpha\beta)(\gamma\delta)}$ must be symmetric with respect to 
$\alpha\leftrightarrow\beta$, $\gamma\leftrightarrow\delta$ or $\alpha\beta\leftrightarrow\gamma\delta$, 
and this gives equation (\ref{eqA:eta}) (see also~\cite{lubensky2003}). 
On the other hand, the rotational viscosity $\eta_{[\alpha\beta][\gamma\delta]}'$ is anti-symmetric with respect to 
$\alpha\leftrightarrow\beta$, $\gamma\leftrightarrow\delta$ but symmetric with respect to $\alpha\beta\leftrightarrow\gamma\delta$.
Finally, the viscosity $\zeta_{\alpha\beta\gamma\delta}$ is symmetric with respect to $\alpha\beta\leftrightarrow\gamma\delta$ but has no further restrictions. 
Note that the notation $\eta_{(\alpha\beta)(\gamma\delta)}(\{\alpha_{i}\})$ and $\eta'_{[\alpha\beta][\gamma\delta]}(\{\alpha_{i}'\})$ in (\ref{eqA:zeta})
mean that $\zeta_{\alpha\beta\gamma\delta}$ has the same structure as these two combined. 
We will use the same notation below for brevity.

A similar use of symmetries gives the form of the reactive couplings $X_{\alpha\beta\gamma}$'s as:
\begin{align}
X_{\alpha(\beta\gamma)}^{\nu}(\{\chi_{i},k_{i}\}) & = \chi_{1}p_{\alpha}\delta_{\beta\gamma}
									   + \chi_{2}(p_{\beta}\delta_{\alpha\gamma}+p_{\gamma}\delta_{\alpha\beta})
									   + \chi_{3}p_{\alpha}p_{\beta}p_{\gamma} \nonumber  \\
									& + k_{1}(\epsilon_{\alpha\beta\nu}p_{\nu}p_{\gamma}+\epsilon_{\alpha\gamma\nu}p_{\nu}p_{\beta}) \, ,\\
X_{\alpha[\beta\gamma]}^{\omega}(\{\chi_{i},k_{i}\}) & = \chi_{1}\epsilon_{\beta\gamma\nu}\epsilon_{\alpha\mu\nu}p_{\mu}
										  + k_{1}\epsilon_{\alpha\beta\gamma}+k_{2}\epsilon_{\beta\gamma\delta}p_{\delta}p_{\alpha} \, ,\\
X_{\alpha\beta\gamma}^{\Omega}(\{\chi_{i},k_{i},\chi_{i}',k_{i}'\}) & = X_{\alpha(\beta\gamma)}^{\nu}(\{k_{i},\chi_{i}\}) + X_{\alpha[\beta\gamma]}^{\omega}(\{k_{i}',\chi_{i}'\}),
\end{align}
where $\{\chi_{i},k_{i},\chi_{i}',k_{i}'\}$ are all functions of $|\vecp|^2$ and $n$. 
Here, $\chi_{i}$'s denote achiral term whereas $k_{i}$'s denote chiral terms. 
Note that the chiral terms in $\boldsymbol{X}^{\Omega}$ are the achiral terms in $\boldsymbol{X}^{\nu}$ and $\boldsymbol{X}^{\omega}$
because, unlike $\boldsymbol{\nu}$, $\boldsymbol{\omega}$ and $\boldsymbol{\Omega}$,
the sign of $\boldsymbol{\nabla}\boldsymbol{\Omega}$ is flipped when $\vecr\rightarrow-\vecr$.

Finally, the off-diagonal viscosities in $\boldsymbol{\mathcal{D}}$ have the general form:
\begin{align}
Y_{(\alpha\beta)[\gamma\delta]}^{\omega}(\{\chi_{i},k_{i}\}) & = \epsilon_{\gamma\delta\mu}X_{\mu(\alpha\beta)}^{\nu}(\{k_{i},\chi_{i}\}) \, ,\\
Y_{(\alpha\beta)\gamma\delta}^{\Omega}(\{\chi_{i},k_{i},k'_{i},&\chi_i^{\Omega},k^{\Omega}\}) = \eta_{(\alpha\beta)(\gamma\delta)}(\{k'_{i}\})
					+ Y_{(\alpha\beta)[\gamma\delta]}^{\omega}(\{k_{i},\chi_{i}\}) \nonumber \\
				     & + k^{\Omega}(\delta_{\alpha\beta}p_{\gamma}p_{\delta}-\delta_{\gamma\delta}p_{\alpha}p_{\beta}) \nonumber \\
 				     & +\chi_1^{\Omega}(\epsilon_{\beta\delta\mu}\delta_{\alpha\gamma}+\epsilon_{\beta\gamma\mu}\delta_{\alpha\delta}
 				     +\epsilon_{\alpha\delta\mu}\delta_{\beta\gamma}+\epsilon_{\alpha\gamma\mu}\delta_{\beta\delta})p_{\mu} \nonumber \\ &+\chi_2^{\Omega}(\epsilon_{\beta\delta\mu}p_{\alpha}p_{\gamma}+\epsilon_{\beta\gamma\mu}p_{\alpha}p_{\delta}
 				     +\epsilon_{\alpha\delta\mu}p_{\beta}p_{\gamma}+\epsilon_{\alpha\gamma\mu}p_{\beta}p_{\delta})p_{\mu} \, ,\\
Z_{[\alpha\beta]\gamma\delta}(\{\chi_{i},k_{i}\}) & =\epsilon_{\alpha\beta\mu}X_{\mu\gamma\delta}^{\Omega}(\{k_{i},\chi_{i}\}).
\end{align}

The general forms of the constitutive relations determined above are quite cumbersome so we now make the following approximations. 
First we ignore terms of order $\vecn\vecn\boldsymbol{u}$.
In particular we ignore $\boldsymbol{\nabla}\boldsymbol{\Omega}$ and consequently $\boldsymbol{C}^{\ell}$ and $\boldsymbol{C}^{\sigma,k}$. 
After elimination of the angular momentum equation, these terms can contribute only to higher order gradient terms than those usually retained in hydrodynamic theories. 
We further ignore all the chiral terms $k_{i}$'s. 
In other words, the source of chirality only comes from the activity, and not from passive microstructure. This is not the most general case (see the closing paragraph of Section \ref{conclusion}) but is suggested by some experiments on bacteria~\cite{zhou}. 
With this approximation we have:
\begin{align}
X_{\alpha(\beta\gamma)}^\nu & = \chi_1^\nu p_\alpha\delta_{\beta\gamma}
						   + \chi_2^\nu(p_{\beta}\delta_{\alpha\gamma} + p_{\gamma}\delta_{\alpha\beta})
						   + \chi_3^\nu p_{\alpha}p_{\beta}p_{\gamma} \label{eqA:Xnu} \, ,\\
X_{\alpha[\beta\gamma]}^\omega & =\chi^\omega\epsilon_{\beta\gamma\nu}\epsilon_{\alpha\mu\nu}p_{\mu} \label{eqA:Xomega} \, ,\\
Y_{(\alpha\beta)[\gamma\delta]}^\omega &= \chi^Y \epsilon_{\gamma\delta\mu}(\epsilon_{\mu\alpha\nu}p_\nu p_\beta + \epsilon_{\mu\beta\nu}p_\nu p_\alpha)\, , \label{eqA:XY}
\end{align}
while other coefficients are ignored.

Substituting (\ref{eqA:Xnu}-\ref{eqA:XY}) into (\ref{eqA:flux-force}),
we can recover the equation for $\vecp$ in the main text (\ref{eq:pdot}) by mapping,
$2\chi^\omega=\xi_{1}$, $2\chi_{2}^\nu=\xi_{0}$, and $\chi_{3}^\nu=\xi_{2}$.
Similarly, equations (\ref{eq:sigmad}) and (\ref{eq:sigmar}) in the main text can be obtained by identifying $4\chi^Y=\eta_c$.
Note that  $\chi_1^\nu$ only contributes to the isotropic stress (the pressure), which is a Lagrange multiplier that enforces incompressibility. 
This contribution can thus be neglected.

\section{Derivation of `spurious drift' terms in the stochastic model \label{app-drift}}

Consider the following set of equations of motion [\emph{cf.} equation (\ref{eqA:flux-force})]:
\begin{equation}
\mathcal{J}_{i}(\vecr,t)=\int d\vecr'[\mathcal{D}+\mathcal{R}]_{ij}(\vecr,\vecr';t)\mathcal{F}_{j}(\vecr',t),\label{eqC:deterministic}
\end{equation}
where $\mathcal{J}_{i} = \D \psi_i / \D t$ is the generalized flux and $\mathcal{F}_{i} = -\delta F\{\psi_i\} / \delta \psi_i$ is the generalized force ($F$ is the free-energy). 
$\boldsymbol{\mathcal{D}}$ is the matrix of dissipative coefficients and $\boldsymbol{\mathcal{R}}$ is the one of reactive coefficients.
Note that the time derivative in $\mathcal{J}$ is not partial, {\it i.e.}, it is a convective and co-rotational derivative; 
for example, if $\psi_{i}$ is a vector, $\D \boldsymbol{\psi}_i /\D t=\dot\psi_i + \vecn\cdot(\psi_i\vecu)+\boldsymbol{\Omega}\cdot\boldsymbol{\psi}_i $, 
and  if $\psi_i$ is scalar, $\D \psi_i/ \D t = \dot\psi_i + \vecn\cdot(\psi_i\vecu)$. 

For the passive limit considered here, when we add noise to (\ref{eqC:deterministic}), 
we must also add a drift term, if one is needed, to ensure that the average total rate of entropy production, $\left\langle \dot{S}_{\text{tot}}\right\rangle$, is zero in steady state ~\cite{EPR}, where the probability must be given by the Boltzmann distribution $P\propto e^{-F/k_{B}T}$.
Assuming the It\^{o} convention, as we chose to do in the main text, the resulting Langevin equation is
\begin{equation}
\mathcal{J}_{i}(\vecr,t)=\int d\vecr'(\mathcal{D}+\mathcal{R})_{ij}(\vecr,\vecr';t)\mathcal{F}_{j}(\vecr',t)+\varUpsilon_{i}(\vecr,t)+\underbrace{k_{B}T\int d\vecr'\frac{\delta(\mathcal{D+\mathcal{R}})_{ij}(\vecr,\vecr';t)}{\delta\psi_{j}(\vecr',t)}}_{\mathcal{J}_{i}^{\text{spurious}}} \, ,\label{eqC:Langevin}
\end{equation}
where $\varUpsilon_{i}$ is a Gaussian noise with variance 
\begin{equation}
\left\langle \varUpsilon_i(\vecr,t)\varUpsilon_{j}(\vecr',t')\right\rangle = 2k_BT\mathcal{D}_{ij}(\vecr,\vecr';t)\delta(t-t')\, ,
\end{equation}
and the last term in (\ref{eqC:Langevin}) is the so-called `spurious' drift~\cite{chaikin,lau,basu,dadhichi}. 
It is, of course, not spurious, except in the sense that it is absent from the the noiseless equations despite being a deterministic contribution (albeit proportional to the noise variance via $\kbt$ and varying in form depending on the interpretation rule for the Langevin equation \cite{lau}).

Consider now the effective polar hydrodynamics after elimination of angular momentum (see Section~\ref{elimination} in the main text). 
The effective free energy is
\begin{equation}
F[\vecp,\vecu]=\int d\vecr\left\{ \frac{1}{2}\rho u^{2}+\mathbb{F}_{0}(\vecp,\nabla\vecp)\right\} \, ,
\end{equation}
and the Langevin equations are:
\begin{equation}
\underbrace{\left( \begin{array}{c}
			   \frac{Dp_{\alpha}}{Dt}+\omega_{\alpha\beta}p_{\beta} \\
			    \sigma_{\alpha\beta}^{s,k} \end{array}\right)}_{\mathcal{J}_{i}}
= \underbrace{\left(\begin{array}{cc}
		  \bar{\gamma}_{\alpha\pi} & \bar{\xi}_{\alpha(\gamma\delta)} \\
		-\bar{\xi}_{\pi(\alpha\beta)} & \bar{\eta}_{\alpha\beta\gamma\delta} \end{array}\right)}_{\mathcal{D}_{ij}+\mathcal{R}_{ij}}
   \underbrace{\left(\begin{array}{c}
			-h_{\pi} \\
			 \nu_{\gamma\delta} \end{array}\right)}_{\mathcal{F}_{j}}
+ \underbrace{\left(\begin{array}{c}
			\varLambda_{\alpha} \\
			\varSigma_{\alpha\beta} \end{array}\right)}_{\text{noise}}
+ \underbrace{\left(\begin{array}{c}
			\mathcal{J}_{1,\alpha}^{\text{spurious}} \\
			\mathcal{J}_{2,\alpha\beta}^{\text{spurious}} \end{array}\right)}_{\text{spurious drift}}\, .
\end{equation}
Here we identify $\boldsymbol{\psi}_{1}=\vecp$, where both $\boldsymbol{\mathcal{D}}$ and $\boldsymbol{\mathcal{R}}$ solely depend on $\vecp$. 
The spurious drift for the $\vecp$-dynamics is then
\begin{equation}
\mathcal{J}_{1,\alpha}^{\text{spurious}} =  k_{B}T\int d\vecr'\frac{\delta(\mathcal{D+\mathcal{R}})_{11}(\vecr,\vecr';t)}{\delta\psi_{1}(\vecr',t)}
							    = k_{B}T\int d\vecr'\delta(\vecr-\vecr')\frac{\delta\bar{\gamma}_{\alpha\beta}(\vecr,t)}{\delta p_{\beta}(\vecr',t)}
\end{equation}
In general, we can write $\bar{\gamma}_{\alpha\beta}=\bar{g}_{1}(|\vecp|^{2})\delta_{\alpha\beta}+\bar{g}_{2}(|\vecp|^{2})p_{\alpha}p_{\beta}$, yielding, 
\begin{equation}
\mathcal{J}_{1,\alpha}^{\text{spurious}}= 2 k_{B}T\delta(\mathbf{0})\Big[\bar{g}_{1}'(|\vecp|^{2})+\bar{g}_{2}'(|\vecp|^{2})|\vecp|^{2}+\bar{g}_{2}(|\vecp|^{2})\Big]p_{\alpha}\, ,
\end{equation}
where ${\cal O}'(x) = \D {\cal O}/ \D x$. 
Note that for any local $(\boldsymbol{\mathcal{D}}+\boldsymbol{\mathcal{R}})$,
we get a divergent prefactor of $\delta(\boldsymbol{0})$ whose meaning in this context is well understood as $1/a^3$, where $a$ is the lattice size~\cite{sancho,sancho2003}. 

Next, for the Navier-Stokes equation, the spurious drift (together with the noise term $\varSigma_{\alpha\beta}$) is added to the stress tensor:
\begin{equation}
\mathcal{J}_{2,\alpha\beta}^{\text{spurious} } = k_BT\int d\vecr'\frac{\delta(\mathcal{D+\mathcal{R}})_{21}(\vecr,\vecr';t)}{\delta\psi_{1}(\vecr',t)}
								     = -k_BT\int d\vecr'\delta(\vecr-\vecr')\frac{\delta\bar{\xi}_{\pi(\alpha\beta)}(\vecr,t)}{\delta p_{\pi}(\vecr',t)}.
\end{equation}
Using the general form for 
$\bar{\xi}_{\pi(\alpha\beta)}=\frac{1}{2}\bar{\xi}_{0}(|\vecp|^{2})(p_{\alpha}\delta_{\pi\beta}+p_{\beta}\delta_{\pi\alpha})+\bar{\xi}_{2}(|\vecp|^{2})p_{\pi}p_{\alpha}p_{\beta}$,
we get
\begin{equation}
\mathcal{J}_{2,\alpha\beta}^{\text{spurious}} = -k_{B}T\delta(\mathbf{0})\Big\{\bar{\xi}_{0}(|\vecp|^{2})\delta_{\alpha\beta}
													+[2\bar{\xi}_{0}'(|\vecp|^{2})
													+2\bar{\xi}_{2}'(|\vecp|^{2})|\vecp|^{2}
													+3\bar{\xi}_{2}(|\vecp|^{2})]p_{\alpha}p_{\beta}\Big\}
\end{equation}
The second term contains the dyadic $p_\alpha p_\beta$, which is of the form usually considered for active stress in polar and nematic liquid crystals~\cite{elsen2015,hatwalne}. Possible implications for numerical studies are discussed  after Eq.~\eqref{eq:udot-noise} in the main text.

\end{document}